\documentclass[final,a4paper]{article}
\usepackage{amsmath,amssymb,amsfonts,amsthm}
\usepackage[mathscr]{eucal}
\usepackage{times}
\renewcommand{\gg}{{\mathfrak g}}
\newcommand{\hh}{{\mathfrak h}}
\renewcommand{\ggg}{\hat{\gg}}
\newcommand{\Cbb}{{\mathbb C}}
\newcommand{\Zbb}{{\mathbb Z}}

\newcommand{\G}{{\mathcal G}}
\newcommand{\M}{{\mathcal M}}
\newcommand{\s}{{\mathbf s}}
\newcommand{\sw}{{{\mathbf s}_{w}}}
\newcommand{\Wg}{{\mathcal W}\left( \gg \right)}
\newcommand{\Wgz}{{\mathcal W}\left( \gg_0 \right)}
\newcommand{\Ww}{{\mathscr W}}
\newcommand{\Hw}{{\mathcal H}^{[w]}}
\newcommand{\pb}[2]{\left\{ #1,#2 \right\}}
\newcommand{\spa}{\mspace{9mu}}
\newcommand{\remove}[1]{}
\newcommand{\N}{{\mathcal N}}
\newcommand{\f}[2]{\frac{#1}{#2}}
\newcommand{\e}{\epsilon}
\newcommand{\Del}{\Delta}
\newcommand{\Oe}{{\mathcal O}\left(\e\right)}
\newcommand{\dl}{\delta}
\allowdisplaybreaks[4]
\DeclareMathOperator{\ad}{ad}
\DeclareMathOperator{\Ker}{Ker}
\DeclareMathOperator{\Img}{Im}
\DeclareMathOperator{\Tr}{Tr}
\DeclareMathOperator{\ord}{ord}
\renewcommand{\Pr}{{\mathop {\mathsf{Pr}}}}
\theoremstyle{plain}
\newtheorem{theorem}{Theorem}
\newtheorem{lemma}{Lemma}
\newtheorem{corollary}{Corollary}
\newtheorem{proposition}{Proposition}
\newtheorem{definition}{Definition}
\theoremstyle{definition}
\newtheorem{remark}{Remark}
\numberwithin{equation}{section}
\numberwithin{definition}{section}
\numberwithin{lemma}{section}
\numberwithin{theorem}{section}
\numberwithin{proposition}{section}
\numberwithin{corollary}{section}
\numberwithin{remark}{section}
\title{Solutions to WDVV from generalized Drinfeld-Sokolov hierarchies}
\author{Oleksandr Pavlyk\thanks{
Bogolyubov Institute for Theoretical Physics, Metrologichna str. 14b, Kiev.
}}
\begin{document}
\maketitle
\begin{abstract}
 The dispersionless limit of generalized Drinfeld-Sokolov hierarchies
 associated to primitive regular conjugacy class of Weyl group $\Wg$
 is discussed.  The map from these generalized Drinfeld - Sokolov
 hierarchies to algebraic solutions to WDVV equations has been
 constructed. Example of $\gg=D_4$ and $[w]=D_4(a_1)$ is considered in
 details and  corresponding Frobenius structure is found.
\end{abstract}

\section{Introduction and discussion}
 Equations of Witten, Dijkgraaf, Verlinde and Verlinde~\cite{WDVV}
\begin{equation}
\begin{split}
   c_{ijk}(t) \eta^{kl} c_{lmn}(t) = c_{njk}(t) \eta^{kl} c_{lmi}(t)
   \,, \quad \text{where} \quad
   c_{ijk}(t) = \partial_i \partial_j \partial_k F(t) \\
   \eta^{-1}_{mn} = \partial_1 \partial_m \partial_n F(t) \text{ --
   constant,} \qquad 
  \sum_{i=1}^n d_i t_i \partial_i F(t) = \left(3-d \right) F(t)
\end{split}
\label{eq:WDVV}
\end{equation}
describe deformations of two dimensional topological conformal quantum
field theory.  They have appeared to be intimately related to
integrable systems. This was recognized by
B.~Dubrovin~\cite{DuIntSys}, who showed that each solution to WDVV
equations gives rise to dispersionless bi Hamiltonian integrable
system, that is with Poisson structures of hydrodynamic
type~\cite{DubrovinNovikov} and Hamiltonians not depending on field
derivatives.  This result was extended by one loop
correction~\cite{DZ}, where it was shown that so constructed bi
Hamiltonian structure is always a $\Ww$ algebra, i.e. it contains a
Virasoro subalgebra.

 On the other hand, due to B.~Dubrovin~\cite{DuFlat}, it is possible,
under certain assumption, to recover solution to WDVV equation from bi
Hamiltonian structure of hydrodynamic type
\begin{equation*}
  \pb{u^i(x)}{u^j(y)} = g^{ij}\left(u(x)\right)
  \delta'\left(x-y\right) + \Gamma^{ij}_k\left(u(x)\right) u'_k(x) \delta\left(x-y\right).
\end{equation*}
 If matrix $g$ does not degenerate identically, it defines flat
metrics~\cite{DubrovinNovikov} on the target space. 

 Whitham averaging method~\cite{DubrovinNovikov} provides Poisson
structures of this type. This method allows to describe slow modulated
$m$-phase solutions of non linear Hamiltonian system of equations. If
one needs local Poisson structures of hydrodynamic type we
should apply zero phase averaging being simply dispersionless limit.

 The dispersionless limit of Drinfeld - Sokolov hierarchies~\cite{DS},
associated to non twisted affine Lie algebra $\ggg$ was
shown~\cite{Krichever} to provide polynomial solutions to WDVV
equations~\eqref{eq:WDVV}. These solutions, viewed as Frobenius
structures on orbits of Coxeter groups ${\mathcal G}$~\cite{DuCox},
correspond to Weyl group ${\mathcal G}=\Wg$ of underlying simple Lie
algebra $\gg$. They correspond~\cite{DuMa} to the orbit of braid group
on Stocks matrices $S$ passing through ``Coxeter'' point, i.e. $S +
S^{tr}$ defines a Carter graph~\cite{Carter} of Coxeter conjugacy
class in $\Wg$. Orbits given by primitive conjugacy classes were
conjectured~\cite{DuMa} to correspond to algebraic solutions of WDVV
equations~\eqref{eq:WDVV}.

 In this article we address the dispersionless limit of generalized
Drinfeld - Sokolov hierarchies~\cite{gDSh1,gDSh2}, associated with
$(i)$ non twisted affine Lie algebra $\ggg$, $(ii)$ its Heisenberg
subalgebra $\Hw$, corresponding~\cite{KacPeterson} to regular
primitive conjugacy class $[w]$ in $\Wg$, and $(iii)$ regular element
$\Lambda$ from it. The naive limit is shown to exist only for standard
Drinfeld-Sokolov hierarchies, due to failure to parameterize the whole phase
space by densities $\N_i$ of annihilators of the first Poisson
structure. As a consequence, some fields generically evolve fast. We
combine Hamiltonian reduction with Whitham averaging~\cite{Maltsev},
restricting Poisson structures to the subvariety parameterized by
annihilators' densities and then taking the limit $\e \to 0$.  So
constructed dispersionless limit is proved to give rise to algebraic
solution to WDVV equations~\eqref{eq:WDVV}. Monodromy group of the
Frobenius manifold coincides with $\Wg$.

 The paper is organized as follows. In section~\ref{sec:Prel} we
gather, to make the paper self-content, some preliminary information
about gradations of Kac-Moody algebra and about its Heisenberg
subalgebras.  Section~\ref{sec:gDSh} reviews basic facts about
generalized Drinfeld - Sokolov hierarchies. The properties of regular
primitive conjugacy classes, relevant for the paper, are recollected
in section~\ref{sec:regprim}. Section~\ref{sec:typeI} brings together
facts about type I integrable hierarchies with primitive regular
conjugacy class $[w]$ and regular element $\Lambda$. Quite rich
underlying finite dimensional geometry is analysed in section
\ref{sec:FinDimGeom}. Finite group $R$, acting non linearly on Miura
variables, and a ring of $R$-invariant polynomials are studied.  The
dispersionless limit of the hierarchies in consideration is examined
in section~\ref{sec:zerophase}. The non degeneracy of obtained
metrics is proved, their flat coordinates are found and it is shown that
Frobenius structure can be always extracted from this averaged Poisson
structure.  Finally, section~\ref{sec:example} treats the simplest
integrable hierarchy with non Coxeter regular primitive conjugacy
class, i.e. $\gg=D_4$ and $[w]=[D_4(a_1)]$. Corresponding solution to
WDVV equations is found and finite group $R$ is analysed.

\section{Preliminaries} \label{sec:Prel}
%\paragraph{Gradations.}
 Let $\gg$ be a simple Lie algebra of rank $r$, and $\ggg=\gg \otimes
\Cbb \left[ z,z^{-1}\right] \oplus \Cbb d$ -- its affine Lie
algebra. This algebra is well-known to be graded, with gradations
being in one-to-one correspondence~\cite{Kac:book} with finite order
inner automorphisms of $\gg$.
\begin{definition}
 Let $\s=\left(s_0,s_1\dots,s_r \right)$ -- a sequence of non negative
relatively prime integers. Let $N_\s = \sum_{i=1}^r k_i s_i$, where $k_i$ --
are Kac labels (such that $\alpha_{\text{max}} = \sum_{i=1}^r k_i \alpha_i$ and $k_0=1$).
We then define the finite order automorphisms $\sigma$ of $\gg$ in some
Cartan-Weyl basis by
\begin{equation}
 \sigma \left( H \right) = H, \qquad \sigma \left( E_\alpha \right) =
 e^{2 \pi i \delta_\s \cdot \alpha} E_\alpha\,,
\label{eq:innp1}
\end{equation}
with
\begin{equation}
  \delta_\s = \frac{1}{N_\s} \sum_{k=1}^r \frac{2}{\alpha_i^2} s_k \omega_k
  \,; \qquad \alpha^\vee_i \cdot \omega_j = \delta_{ij} \,.
  \notag
\end{equation}
\end{definition}
\begin{proposition}[\cite{Kac:book}]
 Such automorphisms exhaust the finite order inner automorphisms of $\gg$
up to conjugacy.
\end{proposition}
An inner automorphisms of $\gg$ can be used to define a new $\Zbb$-gradation
of $\ggg$.
\begin{definition}
 A gradation of type $\s$ is defined via derivation $d_\s$
\begin{equation}
  d_\s = N_\s \left( z \frac{d}{dz} + \ad H_{\delta_\s} \right) ,
 \label{eq:sderv}
\end{equation}
where $H_{\delta_\s} \in {\mathfrak h}$ such that
$\left[ H_{\delta_\s}, E_{\alpha_k} \right] =
\left( \alpha_k \cdot \delta_\s \right) E_{\alpha_k} = s_k E_{\alpha_k}/N_\s$.
\end{definition}
Let us denote $\ggg_k\left(\s \right)$ the eigenspace of derivation $d_\s$ with
an eigenvalue $k$:
\begin{equation*}
  \ggg = \bigoplus_{k \in \Zbb} \ggg_k\left(\sw\right), \qquad 
  \left[ d_\s, \ggg_k\left(\sw\right) \right] = k \,
  \ggg_k\left(\sw\right).
\end{equation*}
Homogeneous gradation corresponds then to
$\s_h = \left(1,0,\dots,0\right)$, while principal to
$\s_p = \left(1,1,\dots,1\right)$.
 One can introduce a partial ordering on the set of gradation.
\begin{definition}
 We say that $\s \succcurlyeq \s'$ if $s_i \not= 0$ whenever $s'_i \not= 0$.
\end{definition}

\begin{proposition}[\cite{gDSh1}] \label{prop:partord}
 If $\s \succcurlyeq \s'$ then the following is true
\par (i) $\ggg_0 (\s) \subseteq \ggg_0 (\s')$,
\par (ii) $\ggg_{>0}(\s) \subseteq \ggg_{\geqslant 0} (\s')$ and
       $\ggg_{<0}(\s) \subseteq \ggg_{\leqslant 0} (\s')$,
\par (iii) $\ggg_{>0}(\s') \subseteq \ggg_{>0}(\s)$ and
       $\ggg_{<0}(\s') \subseteq \ggg_{<0}(\s)$.
\end{proposition}

\begin{theorem}[\cite{KacPeterson}]
 Given a Kac-Moody algebra $\ggg$, its maximal nilpotent subalgebras
are called Heisenberg subalgebras. Heisenberg subalgebras are in
one-to-one correspondence with conjugacy classes of Weyl group $\Wg$
of $\gg$.
\end{theorem}
 Hence, they will be  denoted by $\Hw$. Let us recall the essentials of
their construction. 

 Let $w$ be a representative of a given conjugacy class $[w]$ acting
naturally on $\hh$ in some Cartan-Weyl basis marked by a prime to
distinguish it from another basis~\eqref{eq:innp1} in use. Its action
on Cartan subagebra $\hh$ is known to be inner in $\gg$.  Then, let $Y
\in \gg$ be such an element that $w = \exp \left(2 \pi i \ad Y \right)_{\vert
\hh}$. Let $\hat{w}$ be an obvious extension of $w$ on the whole
algebra $\gg$:
\begin{equation}
  \hat{w} H'_\alpha = H'_{w(\alpha)}\,, \qquad 
  \hat{w} E'_{\alpha} = \psi_\alpha E'_{w(\alpha)}\,,
 \label{eq:PrimedExt}
\end{equation}
where $\psi_\alpha$ are structure constant compatible factors taking values
in $S^1$ and may be chosen to be real.  It can be
shown~\cite{KacPeterson} that the order of this extension
$N_{\hat{w}}$ may only be either the same as or twice as the order of
$w$. The following construction does not, however, depend on this
ambiguity.

Let $a \in \gg$ be a representative of some orbit of $\hat{w}$. We
write $a = \sum_\xi a_\xi$ as a sum of $\hat{w}$ eigenvectors $a_\xi$ with
distinct eigenvalues $\xi = \exp\left[ -2 \pi i \frac{k}{N_{\hat{w}}}
\right]$ and put $\Tilde{a}(k) \in \ggg$ for any $k\in \Zbb$ to be 
\begin{equation}
    \Tilde{a}(k) = \exp\left[ i \varphi \frac{k}{ N_{\hat{w}}} \right] 
    \exp\left[ - i \varphi \ad Y \right]  a_\xi + \delta_{k,0} \left(a
    \vert Y \right) c \,, 
 \label{eq:HeisenbergDef}
\end{equation}
where $c$ is the central element of Kac-Moody algebra. It is obvious
that $\Tilde{a}(k)$ is well defined on $S^1$ and hence depends on
$\varphi$ through $z= \exp( i\varphi)$ only.  It verifies commutation
relations of Kac-Moody algebra:
\begin{equation*}
   \left[ \Tilde{a}(k), \Tilde{b}(l) \right] =
   \widetilde{\left[a,b\right]}\left(k+l\right) + k \delta_{k,-l}
   \left(a \vert b \right) c \,.
\end{equation*}
 The image of $\hh$ under this map is a subalgebra of $\ggg$ isomorphic to
an infinite dimensional Heisenberg algebra:
\begin{equation*}
   \left[ \Tilde{h}_1(k), \Tilde{h}_2(l) \right] =
    k \delta_{k,-l}
   \left(h_1 \vert h_2 \right) c \,.
\end{equation*}
 Factorizing out the center ${\mathfrak z} = \Cbb c$ of Kac-Moody
algebra $\ggg$, the Heisenberg subalgebras become maximal commutative
ones.  A practical realization of this construction faces the
difficulty to find $Y$ explicitly.

Associated to each Heisenberg subalgebra $\Hw$ there is a distinguished
gradation $\sw$ with the property that $\Hw$ admits $\sw$ grading~\cite{gDSh1}:
\begin{equation*}
 \Hw = \bigoplus_{k \in E} \Hw_k \left(\sw \right) \,,
\end{equation*}
with
\begin{equation}
\begin{split}
 E &= \left\{ k + N_{\hat{w}} \Zbb \, \vert k \in I(w) \right\} ,\\
 I (w) &= \left\{ k_n \in \Zbb_+ \,, n=1,\dots,r \, \vert \, 0 \leqslant k_1 \leqslant k_2
 \leqslant \dots \leqslant k_r < N_{\hat{w}} \,, \right. \\ 
 & \mspace{220mu} \left.  e^{2 \pi i k_n/N_{\hat{w}}} \text{ -- eigenvalue of } w \right\} .
\end{split}
 \label{eq:Iw}
\end{equation}
It is clear, that $\sw$ is the gradation corresponding to automorphism
$w$ used to construct $\Hw$. Indeed, let us choose $Y\in \gg$ so that
spectrum of $N_{\hat{w}} \ad Y$ on $\hh$ coincides with $I\left(w\right)$. Then 
\begin{equation*}
  d_\s = N_{\hat{w}} \left( z \frac{d}{dz} + \ad Y \right) , \qquad d_\s
  \left( \Tilde{h}\left(k \right) \right) = k \, \Tilde{h}\left(k
  \right) .
\end{equation*}

\begin{definition}
 An element $\Lambda \in \ggg$ is called \emph{semisimple} if
 $\ggg = \Ker \left( \ad \Lambda \right) \oplus \Img \left( \ad \Lambda
 \right)$ and $\Ker \left( \ad \Lambda \right) \cap \Img \left( \ad \Lambda
 \right) = \emptyset$.
\end{definition}

\begin{definition}
 A semisimple element $\Lambda \in \Hw$ is called \emph{regular} if
 $\Ker \left( \ad \Lambda \right)  = \Hw$.
\end{definition}

\section{Generalized integrable hierarchies} \label{sec:gDSh}

 Let $[w]$ be some conjugacy class in $\Wg$, $\sw$ -- corresponding
gradation.  Pick up any $\s \succcurlyeq \sw$ and some constant
semisimple element $\Lambda \in \Hw$ of certain $\sw$ grade $i \in
E_+$. Let us define the matrix Lax operator~\cite{gDSh1}
\begin{equation}
 {\mathcal L} = \partial_x + \Lambda + q(x) \,,
 \label{eq:LaxOp}
\end{equation}
 where $q(x)$ takes value in
\begin{equation}
 Q\left(i\right) = \ggg_{\geqslant 0} \left( \s \right) \cap
  \ggg_{<i} \left( \sw \right) .
 \label{eq:Q}
\end{equation}

\begin{proposition}[\cite{gDSh1,DS,Wilson}]
 There exist a unique formal series
\begin{equation}
T(q) \in \Img \left(\ad \Lambda \right) \cap
 \ggg_{<0}\left(\sw\right)
 \notag
\end{equation}
 that
\begin{equation}
 L = \exp\left[ - \ad T \right] {\mathcal L} = \partial_x +\Lambda +
  h\left(q\right) , \qquad
 h(q) =  \sum_{k<i} h_k(q) 
\label{eq:L0}
\end{equation}
 where $h\left(q\right) \in \Hw_{<i}\left(\sw\right)$. Both
 $P^{(\sw)}_{k} T(q)$ and $P^{(\sw)}_{k} h(q)$ are polynomials in $q(x)$
 and its $x$-derivatives for any $k \in \Zbb$.
\end{proposition}
 There is a gauge symmetry acting on \eqref{eq:LaxOp} and mapping $Q(i)$
 to $Q(i)$:
\begin{equation}
 {\mathcal L } \mapsto \exp \left[ - \ad n(x) \right] {\mathcal L} \,,
\label{eq:gaugetr}
\end{equation}
where $n(x)$ takes values in 
\begin{equation}
  P = \ggg_0 \left( \s \right) \cap \ggg_{<0} \left(\sw \right).
  \label{eq:P}
\end{equation}

\begin{proposition}[\cite{DS,Fernandez-Pousa:1995sj}]\label{prop:ring}
  If $\Lambda$ is chosen so that
\begin{equation}
 \Ker \left( \ad \Lambda \right) \cap P = \emptyset \qquad \Rightarrow
 \qquad \ad \Lambda : P \hookrightarrow Q\,,
 \label{eq:GaugeFixCond}
\end{equation}
 then elements of $Q^{\text{can}}\left(i\right)^\ast$, where
\begin{equation}
  Q(i) = Q^{\text{can}}(i) \oplus \left[ \Lambda , P \right] ,
  \label{eq:Qcan}
\end{equation}
are generators of the ring of gauge invariant polynomials in $q(x)$ and
its derivatives; \eqref{eq:Qcan} holds as equality of vector
spaces. Lax operator $L$ in \eqref{eq:L0} is gauge invariant.
\end{proposition}

\begin{definition}
 Denote by ${\mathcal M}$ the phase space, spanned by the elements of
$Q^{can}\left(i\right)^\ast$ and by ${\mathcal F}\left({\mathcal M}
\right)$ the space of functionals of the type
\begin{equation}
  \varphi \left[q \right] = \int_{S^1} dx \, f\left(x,q,q', \dots
  \right) .
\notag
\end{equation}
\end{definition}

\begin{proposition}[\cite{gDSh1}]\label{prop:dimofM}
 Dimension of phase space ${\mathcal M}$ is
 independent on auxiliary gradation $\s$ and equals to
\begin{equation}
 \dim {\mathcal M} = \dim Q^{\text{can}}(i)^\ast = \sum_{k=0}^{i-1} \dim \ggg_k\left(\sw\right)
 \label{eq:dimM}
\end{equation}
\end{proposition}
\begin{remark}
 $\dim {\mathcal M} \geqslant \dim \ggg_0(\sw) \geqslant \dim \ggg_0(\s_p) =
 r$.
\end{remark}

 Condition \eqref{eq:GaugeFixCond} is automatically
satisfied~\cite{Fernandez-Pousa:1995sj} if $\Lambda$ is chosen to be
regular. 

\begin{theorem}[\cite{gDSh1}]\label{thm:commofflows}
 Given $b \in \Hw_{>0}\left(\sw\right)$, let
 ${\mathcal A}\left(b\right) = \exp\left[\ad T \right] b$.
  One defines two sets of time flows
\begin{equation}
 \frac{\partial}{\partial t_b} {\mathcal L} = \left[
 P^{(\s)}_{\geqslant 0} {\mathcal
 A}\left(b\right) , {\mathcal L} \right]      \qquad
 \frac{\partial}{\partial t_b'} {\mathcal L} = \left[
 P^{(\sw)}_{\geqslant 0} {\mathcal
 A}\left(b\right) , {\mathcal L} \right]
\label{eq:flows}
\end{equation}
being commutative within each set
\begin{equation}
\forall \, b_1,b_2 \in \Hw \qquad
\left[ \frac{\partial}{\partial t_{b_1}} ,
\frac{\partial}{\partial t_{b_2}}  \right] =0 \,, \qquad
\left[ \frac{\partial}{\partial t'_{b_1}} ,
\frac{\partial}{\partial t'_{b_2}}  \right] =0 \,.
\notag
\end{equation}
 Both time flows \eqref{eq:flows} preserve the
 phase space of gauge invariants ${\mathcal M}$. They coincide there and
 retain their commutativity property.
\end{theorem}

\begin{definition}
 Introduce the pairing on the space of functions in
 $C^\infty\left(S^1,\ggg\right)$
\begin{equation}
  \langle A , B \rangle = \int_{S^1} dx  \sum_{k \in \Zbb} \eta\left(
   P^{(\s)}_k A(x), P^{(\s)}_{-k} B(x) \right)_{\ggg_0(\s)}
\end{equation}
 for some gradation $\s$ and Killing form $\eta$ on $\ggg_0(\s)$.
\end{definition}
If $\eta_{\ggg_0(\s)}$ is properly normalized, the pairing does not
depend on the gradation $\s$ chosen. We will assume this normalization
chosen further on.

\begin{proposition}[\cite{DS,gDSh1}]
 Gauge invariant functionals $H_b \left[q \right] = \langle
 b, h(q) \rangle$ are integrals of flows \eqref{eq:flows}.
\end{proposition}
\begin{remark}
 $H_b \equiv 0$ if $\deg_\sw \left( b \right) \leqslant -i$. 
\end{remark}
\begin{definition} \label{def:grad}
 For any $\varphi \in {\mathcal M}$ define gradient $d_q \varphi \in
 \ggg_{\leqslant 0} \left(\s\right) / \ggg_{<-i}\left(\sw\right)$ by
\begin{equation}
 \frac{d}{d \varepsilon} \varphi \left[q+\varepsilon
 r\right]_{\vert_{\varepsilon=0}} = \langle r, d_q \varphi \rangle
 \qquad \forall\, r \in C^\infty \left(S^1,Q(i)\right)
\notag
\end{equation}
\end{definition}

\begin{theorem}[\cite{gDSh2}]\label{thm:BiHam}
 Let $\s=\s_h$, then 
 \par{(i)} there is a one parameter family of Hamiltonian
 structures on the gauge equivalence classes of the generalized
 Drinfeld-Sokolov hierarchy given by
\begin{equation}
  \pb{\varphi}{\psi}_\lambda =
  \langle \Lambda +q , \left[ d_q \varphi, d_q \psi \right]_{R^{(\s)}} \rangle
  - \langle d_q \varphi, \left( d_q \psi \right)' \rangle
\label{eq:BiHamStr}
\end{equation}
where $R^{(\s)}=\left( P^{(\s)}_0 - P^{(\s)}_{<0} \right)/2 - \lambda
/z$. Expanding in powers of $\lambda$, $\pb{\spa}{\spa}_\lambda =
\pb{\spa}{\spa}_2+ \lambda
\pb{\spa}{\spa}_1$, we obtain two coordinated Hamiltonian
structures on ${\mathcal M}$:
\begin{eqnarray}
  \pb{\varphi}{\psi}_1 &=& - \langle d_q \varphi, z^{-1} \left[ d_q \psi,
  {\mathcal L} \right] \rangle \,, \nonumber \\
  \pb{\varphi}{\psi}_2 &=& \langle \Lambda +q , \left[ d_q \varphi,
  d_q \psi \right]_{R} \rangle - \langle d_q \varphi, \left( d_q
  \psi \right)' \rangle \,,\nonumber
\end{eqnarray}
 where $R=\left( P^{(\s)}_0 - P^{(\s)}_{<0} \right)/2$. Under the time
 evolution in the coordinate $t_b$, the following recursion relation
 holds:
\begin{equation}
 \frac{\partial \varphi}{\partial t_b} = \pb{\varphi}{H_{zb}}_1 =
 \pb{\varphi}{H_{b}}_2 \,.
\label{eq:HamRec}
\end{equation}
\par{(ii)} Hamiltonians $H_b$ with $-i < \deg_\sw b < 0$ are
Casimirs of~\eqref{eq:BiHamStr}. 
Hamiltonians  $H_b$ with $\deg_\sw
b=0$  are Casimirs of the first bracket only.
\end{theorem}
 Hierarchies~\eqref{eq:HamRec} with $\Lambda$ regular were dubbed in 
ref.~\cite{gDSh1} the hierarchies of type I.
\begin{remark}
The hierarchies with grade one element $\Lambda$ are distinguished by
the fact that pencil of Poisson structures has no local annihilators. 
This is true also for the first structure as long as conjugacy class $[w]$
is chosen to be non degenerate, that is fixing no vector in $\hh$ and
consequently $0 \not\in I(w)$.
\end{remark}

\begin{corollary} \label{cor:IndepHam}
 Hamiltonians $H_b$, with $b \in \Hw$ of positive $\sw$ degree
belonging to $I(w)$, annihilate the first Poisson structure.

 There are $r$ independent annihilators of $\pb{\cdot}{\cdot}_1$, that
yet generate nontrivial flows with respect to $\pb{\cdot}{\cdot}_2$.
\end{corollary}
\begin{proof}
 Choose $b$ such that $0 \leqslant \deg_\sw \left(zb\right) < N_w$. Then $-N_w
 \leqslant \deg_\sw \left(b\right) <0$ and $H_b$ annihilates
 $\pb{\cdot}{\cdot}_2$ or vanishes identically. 

 To prove independence, it is enough to notice that equation \eqref{eq:L0}
 determining $h(q)$, projected on eigenspaces of $d_{\sw}$ reads
\begin{equation*}
    h_k + \left[ \Lambda, T_{k+1} \right] = q_k +
    \text{Polynomial}\left(h_{i<k}, T_{j \leqslant k}, q_{i<k} \right).
\end{equation*}
 Thus, $h_k$ with $k \in I(w)$ are linear in Drinfeld-Sokolov variable
of scaling degree $k+1$ with successive terms involving variables of
lower scaling weights and this proves independence.

\end{proof}

Notice, that $\s=\s_h$ corresponds to the largest gauge group, while
$\s=\sw$ -- to no gauge freedom at all. The former choice leads to generalized
Drinfeld-Sokolov hierarchies, while the latter -- to modified generalized
Drinfeld-Sokolov hierarchies.

\begin{theorem}[\cite{gDSh2}]\footnote{In~\cite{gDSh2} it was proved
for partially modified hierarchies, that is for any $\s$ such that $ \s_h
\succcurlyeq \s \succcurlyeq \sw$.}
 Let ${\mathcal M}_{m}$ be the phase space of modified hierarchy, and
 $\pb{\spa}{\spa}_m$ -- its second Poisson bracket:
\begin{equation}
 \pb{\varphi}{\psi}_m = \langle q_m +\Lambda, \left[ d_{q_m} \varphi
 ,d_{q_m} \psi \right]_{R_m} \rangle - \langle d_{q_m} \varphi, \left(
 d_{q_m} \psi \right)' \rangle \,,
 \label{eq:MiuPB}
\end{equation}
 The Miura mapping
 is a (non invertible) Hamiltonian mapping,
\begin{equation}
 \mu : \left( {\mathcal M}_m, \pb{\spa}{\spa}_m \right) \to
       \left( {\mathcal M}, \pb{\spa}{\spa}_2 \right) , \label{eq:MiuraMap}
\notag
\end{equation}
such that it defines a reduction of the dynamical equations of the
hierarchy to those of the modified hierarchy.
\end{theorem}

Drinfeld-Sokolov hierarchies~\cite{DS} are recovered picking up
Coxeter conjugacy class $[w_c]$ with representative $w_c =
\prod_{k=1}^r r_{\alpha_k}$, associated principal gradation
($\sw=\s_p$) and regular element $\Lambda= z
E_{-\alpha_{max}}+\sum_{k=1}^r E_{\alpha_k}$ of grade one.

\section{Regular primitive conjugacy classes of $\Wg$}\label{sec:regprim}

Conjugacy classes in Weyl group $\Wg$ of simple Lie algebra $\gg$ were
uniformly described by R.~Carter~\cite{Carter} in terms of Carter
graphs.
\begin{definition}
 Conjugacy class $[w] \subset \Wg$ is called \emph{non degenerate} if
 $\det(1-w) \ne 0$ and is called \emph{primitive} if $\det(1-w) =
 \det(1-w_c)=\det(K)$, $K$ -- being the Cartan matrix.
\end{definition}
Primitive conjugacy classes are distinguished as they have no
representative in any proper Weyl subgroup $W' \subset \Wg$. All
primitive conjugacy classes were listed by R.~Carter~\cite{Carter}
together with their characteristic polynomials
$\det(t-w)$. 
\begin{definition}[\cite{DelducFeher}] \label{def:Reg}
  Conjugacy class $[w] \subset \Wg$ is called \emph{regular} if
  associated Heisenberg subalgebra $\Hw$ admits regular element
  $\Lambda$.
\end{definition}
Regular conjugacy classes were elegantly studied by
T.~Springer in ref.~\cite{Springer}, though another definition of regularity
was used.
\begin{definition}[\cite{Springer}]\label{def:Spr}
 If $G$ is a finite reflection group in a finite dimensional vector
 space $V$, then $v \in V$ is called \emph{regular} if no nonidentity element
 of $G$ fixes $v$. 

 An element $g \in G$ is \emph{regular} if it has a regular eigenvector.
\end{definition}
Two definitions were shown~\cite{DelducFeher}(see also sec.~9 of
\cite{Springer}) to be equivalent. The authors of the
reference~\cite{DelducFeher} studied generalized hierarchies
associated with classical Lie algebras $\gg$ and regular conjugacy
classes. They synthesized available information on regular primitive
conjugacy classes.
\begin{theorem}[\cite{Springer}]\label{th:Springer1}
Let $[w] \in \Wg$ be a regular conjugacy class of order $N$, i.e.
$\forall w \in[w], \, w^N=1$. Then eigenvalues of $w$ are $\exp\left(
2 \pi i {k_n}/{N}\right)$, where $k_n \in I(w_c)$ are exponents. The
elements of the root system $R$ of $\gg$ are permuted by $\forall \, w
\in [w]$ in orbits of length $N$.
\end{theorem}
 So $\{1,N-1\} \subset I(w)$, that implies ${\sw}_0=1$.
\begin{theorem}[\cite{Springer}]\label{thm:Springer2}
 Let $[w]\subset \Wg$ be regular primitive conjugacy class. Let
 $\Lambda=I_+ +z C_{-(N-1)} \in \Hw_1(\sw)$ be grade one regular
 element. Then
 \par{(i)} $\exists I_- \in \gg$ of $\sw$ grade $-1$, such that
 $\rho = N_w \delta_\sw$, $I_+$ and $I_-$ form $sl_2$ subalgebra
\begin{equation}
  \left[ \rho, I_\pm \right] = \pm I_\pm \qquad
  \left[ I_+, I_- \right] = 2 \rho\,;
\label{eq:sl2}
\end{equation}
\par{(ii)} Eigenvalues of $\ad \rho$ on $\gg$ are integers;
\par{(iii)}
\begin{equation}
  \dim \gg(0) = \dim \gg(\pm 1) = \frac{2}{N} \sum_{k=1}^r p_i =
  \frac{ r h}{N} \,, \quad p_i \in I\left(w_c\right)
\label{eq:match}
\end{equation}
\end{theorem}
\begin{corollary}
The extension $\hat w$ of $w$ to the whole algebra $\gg$ is of order $N$.
\end{corollary}
\begin{proof}
 The order of extension does not depend on the basis chosen. In the
 basis of \eqref{eq:innp1} $\hat{w}=\exp\left[2 \pi i \ad \rho
 /N\right]$. Then $\hat{w}^N=1$ as a straightforward consequence of
 statement $(ii)$ of the theorem. 
\end{proof}

Dimensions of $\gg(k)$ may be read off from the character of this
(reducible) representation of $sl_2$ on $\gg$:
\begin{equation}
%\begin{split}
  \chi_{[w]}\left(q\right) = \Tr \left( q^{\ad \rho }\right) = r +
  \sum_{\alpha \in R} q^{\left(\alpha,\rho\right)} =
  \sum_{k=-(N-1)}^{N-1} \dim \gg(k) q^k \,.
%\end{split}
\label{eq:char}
\end{equation}

To determine the $\s$ type of regular primitive conjugacy class we
compare $\chi_{\left[\hat{w}\right]}$ in the basis of \eqref{eq:innp1} 
and in that of \eqref{eq:PrimedExt}:
\begin{equation}
  \Tr w = r + 2 \sum_{\alpha \in R^+} \cos \left( 2 \pi \delta_\sw \cdot
  \alpha \right) = \chi_{[w]}\left(e^{2\pi i / N_w}\right)
\label{eq:cond1}
\end{equation}
  The knowledge of the characteristic polynomials of $[w]$ due
to~\cite{Carter} allows to compute the order $N_w$ of $[w]$. Then one
searches for $\sw$ with $N_w=N$ and ${\sw}_0=1$, satisfying
\eqref{eq:cond1} and \eqref{eq:match}. This and some more (see
sec.~\ref{sec:typeI}) information about regular primitive conjugacy
classes relevant for construction of generalized hierarchies is
collected in appendix~\ref{app:uno}.

\section{Type-I, $i=1$ hierarchies} \label{sec:typeI}

Here and further on we restrict ourselves to hierarchies with a
regular primitive conjugacy class and a regular element $I_+ + z
C_{-(N-1)} = \Lambda \in \Hw$ of $\sw$ grade one. Notice, that
whatever auxiliary gradation $\s$ we choose, $Q(i) \subset \gg$ and we
are working with finite dimensional Lie algebra. Adjoint action of
$\rho$ induces gradation of $\gg$:
\begin{equation*}
   \gg = \bigoplus_{k=-N+1}^{N-1} \gg_k\,.
\end{equation*}
Notice, that $\ggg_0\left(\sw\right)$ coincides with $\gg_0$.

Consider first homogeneous auxiliary gradation $\s=\s_h$.  As follows
from definition~\ref{def:grad} and proposition~\ref{prop:partord}
gradients have no negative $\sw$ grade part: $P^{(\sw)}_{<0} d_q
\varphi = 0 \mod \ggg_{<-1}(\sw)$. So the bi Hamiltonian
structure~\eqref{eq:BiHamStr} simplifies to
\begin{equation}
\begin{split}
 \pb{\varphi}{\psi}_1 &= - \langle d_q \varphi, \left[ d_q \psi ,
 C_{-(N-1)} \right] \rangle \,, \\
 \pb{\varphi}{\psi}_2 &= \langle d_q \varphi, \left[ d_q \psi ,
 \partial_x + I_+ + q \right] \rangle \,.
\end{split}
\label{eq:i1BhSt}
\end{equation}
As we have no $R$ matrix in commutator now, we can recognize in the
second bracket a Kirillov-Poisson bracket corresponding to untwisted
affinization $\ggg_x$ in $x$ of $\gg$. Introducing $J=I_+ +q$ we
obtain
\begin{equation}
  \pb{\varphi}{\psi}_{KM} = \langle d_J \varphi, \left[ d_J\psi , \partial_x
  + J \right] \rangle \,. \label{eq:KMham}
\end{equation}
This fact enabled authors of ref.~\cite{BalFeh} to invent a practical
algorithm to compute Hamiltonian structure of Drinfeld-Sokolov
hierarchies and their construction can be repeated here.  They showed
that the second Poisson structure~\eqref{eq:i1BhSt} may be obtained by
Hamiltonian reduction of~\eqref{eq:KMham} (see also~\cite{Gregorio}).
Let us embed $Q^{can} \subset \ggg_x$ and lift flows on $Q^{can}$ to
flows on $\ggg_x$. One may choose functional $\Psi$ on $\ggg_x$
coinciding with $\psi$ on $Q^{can}$ such that $\Psi$ flows in $\ggg_x$
preserve $Q^{can}$. Then
\begin{equation}
  \delta q = \left[ d_J \Psi_{\vert Q^{can}}, \partial_x +I_+ + q \right]
  = \left[ d_q \psi(x) + r, \partial_x +
  I_+ + q(x) \right] \in Q^{can} \,,
\label{eq:flowsinKM}
\end{equation}
where $\eta\left(r, Q^{can}\right)=0$.  Condition~\eqref{eq:flowsinKM}
determines $r$ uniquely as a function of $q$ and its $x$-derivatives.

The grading of Kac-Moody algebra induces a grading on the Poisson
structure, namely -- the following theorem holds true.
\begin{theorem}[\cite{gDSh2}]
  The dynamical equation~\eqref{eq:flows} of the hierarchy generated by
  the Hamiltonian $H_b$ ( $b \in \Hw_j(\sw)$) with respect to the second
  Hamiltonian structure, is invariant under the following rescaling
\begin{equation}
 x \mapsto \lambda x\,, \qquad t_b \mapsto \lambda^{j} t_b\,, \qquad
 q_{-k} \mapsto \lambda^{-(k+1)} q_{-k} \,,
 \label{eq:scaling}
\end{equation}
where $q_{-k}$ is the component of $q(x)$ with $\sw$ grade $-k$.
\end{theorem}
This is a consequence of existence of Virasoro algebra due to Sugawara
construction.
\begin{theorem}[\cite{Fernandez-Pousa:1995sj,BalFeh}] \label{thm:Walg}
  The second Poisson bracket of~\eqref{eq:i1BhSt} is a $\Ww$ algebra
  associated to $sl_2$ subalgebra~\eqref{eq:sl2} with Virasoro
  generator
\begin{equation}
   w_2(x) = \frac{1}{2} \eta \left( I_+ + q, I_++q \right) + \eta \left( \rho, q' \right) ,
\label{eq:Virgen}
\end{equation}
satisfying Virasoro algebra
\begin{equation}
  \pb{w_2(x)}{w_2(y)} = - \eta \left(\rho,\rho\right)
  \delta'''\left(x-y\right) + 2 w_2(x) \delta'\left(x-y\right) +
  w'_2(x) \delta\left(x-y\right) .
\label{eq:Viral}
\end{equation}
\end{theorem}
 There exists a minimal weight Drinfeld-Sokolov
 gauge~\cite{BalFeh,DelducFeher}
\begin{equation} 
 \ad I_- \left( Q^{can} \right) = 0
\label{eq:MinWeightGauge}
\end{equation}
of conformal primaries:
\begin{equation}
 \pb{w_{k,i}(x)}{w_2(y)}_2 = k w_{k,i}(x) \delta'\left(x-y\right)
  + w'_{k,i}(x)\delta\left(x-y\right) ,
\label{eq:Primaries}
\end{equation}
 where
\begin{equation*}
 Q^{can} \ni q^{can} = w_2 \frac{I_-}{\eta(I_+,I_-)} + \sum w_{k,i}
 F_{k-1,i} \,, \qquad \left[ \rho, F_{k,i} \right] = - k F_{k,i} \,.
\end{equation*}

 Let us denote by $\Pr_w$ a set of scaling weights of fields of the
 $\Ww$-algebra in question. $\Pr_w$ is invariant with respect to scale
 preserving changes of coordinates, so in particular it does not
 depend on the Drinfeld-Sokolov gauge slice chosen.

\begin{proposition} \label{prop:incl}
 $I(w) \subseteq \Pr_w$. Multiplicities of $N_w-1$ in $I(w)$ and in $\Pr_w$
 are equal.
\end{proposition}
\begin{proof}
Let us take densities of Hamiltonians $H_b$, $\deg_\sw b \in
I(w)$. Their scaling degree are correspondingly $\deg_\sw b +1$.
These densities are gauge invariant, as follows from
proposition~\ref{prop:ring}. Due to corollary to the
theorem~\ref{thm:BiHam} they are independent.  Thus they may be taken
as a part of coordinates on ${\mathcal M}$. So $I(w) \subseteq \Pr_w$.
In particular this implies that the multiplicity of $N_w-1$ in $I(w)$ is
less or equal to the one in $\Pr_w$.

However, as was shown in ref.~\cite{Springer}, for every $C \in
\gg(-N_w+1)$ there exist such $I \in \gg(1)$ that $I+C$ is regular in
$\gg$. So we have an opposite inequality of multiplicities. This
completes the proof.  
\end{proof}

\begin{remark} \label{rem:AssumOnShiftVar}
 One can choose $F_{N_w -1,1}=C_{-(N_w-1)}$. Then the first
Poisson structure can be always read off the second by shifting
$w_{N_w,1}$ by $-\lambda$ and taking the linear term.
\end{remark}

 The character~\eqref{eq:char} enables us to determine $\ad \rho$
 eigenspace decomposition of $Q^{can}$, and thus $\Pr_w$.  Indeed,
\begin{equation}
  \left(1-q \right) \chi_{[w]} = \sum_{k=-N+1}^{-1} \dim Q^{can}_k
  q^k + {\mathcal O}\left(q\right) ,
  \label{eq:expans}
\end{equation}
as follows from injectivity~\eqref{eq:GaugeFixCond} of mapping $\ad
I_+ : Q_k \hookrightarrow Q_{k+1}$ for negative $k$ and definition of
$Q^{can}$~\eqref{eq:Qcan}.  Then $\Pr_w$ is the set of positive
numbers $k$ such that $\dim Q_{-k}^{can} \ne 0$, every number occurring
$\dim Q_{-k}^{can}$ times.
\begin{proposition} \label{prop:OnUw}
 Let 
\begin{equation}
{\mathcal U}\left(w\right) = \left\{ 1 \leqslant k_1 \leqslant \dots
\leqslant k_{2n} < N_w-1 \vert k_i \in \Pr_w \, \text{\rm but } \, k_i \not\in
I(w) \right\} ,
\end{equation}
 where $2n = \dim \M -r$.  Then $k_i+k_{2n+1-i} = N_w -1$.
\end{proposition}
\begin{proof}
 Brute force checking of data presented in Appendix
 \ref{app:uno}. We lack more elegant proof so far. 
\end{proof}

 Now, consider modified hierarchy, i.e. $\s=\sw$. Its second Poisson
structure~\eqref{eq:MiuPB} simplifies as well
\begin{equation*}
 \pb{\varphi}{\psi}_m = \langle q_m , \left[ d_{q_m} \varphi, d_{q_m} \psi \right]
 \rangle - \langle d_{q_m} \varphi, \left( d_{q_m} \psi \right)' \rangle \,.
\end{equation*}
The phase space~\eqref{eq:Q} of modified hierarchies $(\gg,\Lambda,\sw)$
reads
\begin{equation}
{\mathcal M}=\left(Q^{m}(1)\right)^\ast=\left(\ggg_0(\sw)\right)^\ast \simeq
 \ggg_0(\sw) \,.
\label{eq:i1Qm}
\end{equation}
 Let $X_i$ be a basis of $\ggg_0(\sw)\subset \gg$ with Grahm matrix
 ${\mathcal K}$ being restricted Killing form and let $q_m = \sum_i
 \nu^i X_i$, then
\begin{equation}
 \pb{\nu^i(x)}{\nu^j(y)}_m = \left({\mathcal K}^{-1}\right)^{ij}
                            \delta'\left(x-y\right) + {f^{ij}}_k
                            \nu^k(x) \delta\left(x-y\right) ,
\label{eq:i1MiuPB}
\end{equation}
where ${f^{ij}}_k$ are structure constants of $\ggg_0(\sw)^\ast$.
Notice, that ${\mathfrak h} \subseteq \ggg_0(\sw)$ due to
proposition~\ref{prop:partord}, and thus matrix ${f^{ij}}_k \nu^k$ is
of corank $r$. 

Given $[w]$ one obtains $\ggg_0(\sw)$ from extended Dynkin diagram
removing nodes with ${\sw}_k \ne 0$ and padding with $u(1)$ to
maintain the rank as follows from~\eqref{eq:sderv}.

We collect all pertinent information about generalized
Drinfeld-Sokolov hierarchies with regular primitive conjugacy class
$[w]$ and regular element $\Lambda \in \Hw$ of $\sw$ grade one in
appendix~\ref{app:uno}.

\section{Finite dimensional geometry behind the hierarchy}
\label{sec:FinDimGeom}

 Let us introduce the dispersion parameter $\epsilon$ by rescaling all
derivatives $\partial \to \e \partial$. Then, the leading term of $\e
\to 0$ expansion  of a
Poisson bracket on loop space  will verify Jacobi identity, thus
furnishing another Poisson structure:
\begin{equation}
   \pb{w^i(x)}{w^j(y)}^{(0)} = \frac{1}{\e} A^{ij}(w)
   \dl\left(x-y\right).  \label{eq:LeadingBr}
\end{equation}
We write upper indices to emphasize the covariant nature of objects.
The leading bracket \eqref{eq:LeadingBr} obviously defines finite
dimensional one and we thus end up with pencil of finite dimensional
Poisson brackets corresponding to $\s=\s_h$  integrable hierarchy:
\begin{equation}
  \overline{\pb{w^i}{w^j}}_\lambda = A^{ij}\left(w\right) - \lambda
  B^{ij} \left(w \right). \label{eq:FDbr}
\end{equation}
Notice, that densities of annihilators of the first Poisson bracket
\eqref{eq:i1BhSt} $\N^a=\N^a\left(w\right)$, where $1 \leqslant a
\leqslant r$ are Casimirs of the first finite bracket:
\begin{equation}
  \overline{\pb{f}{\N^a}}_1 = 0\, \qquad \forall f \,, \label{eq:Obv1}
\end{equation}
and are in involution with respect to the second
\begin{equation}
  \overline{\pb{\N^a}{\N^b}}_2 = 0\,. \label{eq:Obv2}
\end{equation}
Eqs.~\eqref{eq:Obv1} and \eqref{eq:Obv2} show that Poisson tensor
$A-\lambda B$ is of corank $r$ for all $\lambda$ and generic point of
$\M$.

Due to Corollary \ref{cor:IndepHam}, one may choose $w^i=\left( \N^a,
u^A \right)$ as coordinates on ${\mathcal M}$, where $u^A$ are
Drinfeld-Sokolov variables with scaling weights $k_A+1$, $k_A \in
{\mathcal U}\left(w\right)$. 

Assume that the coordinate $\N^r$ is chosen to be linear in
Drinfeld-Sokolov variable $w_{N_w,1}$ (cf. Remark
\ref{rem:AssumOnShiftVar}), and so the pencil \eqref{eq:LeadingBr} can
be obtained by means of shifting of $\N^r$ by $-\lambda$.
\begin{lemma} \label{lem:detBconstant}
 Matrix $B^{AB}\left(u,\N\right)$ is non degenerate with constant
determinant. 
\end{lemma}
\begin{proof}
 As follows from scaling weight grading of Poisson structures 
\begin{equation*}
\deg_{sc}\left( A^{AB} - \lambda B^{AB} \right) = \deg{u^A} + \deg{u^B}-1 \,.
\end{equation*}
Since $\deg_{sc} \lambda = N_w$
\begin{equation*}
  B^{AB} = \begin{cases} 
              0 & \text{ if } \deg_{sc} u^A u^B < N_w \cr
             \text{const. } & \text{ if } \deg_{sc} u^A u^B = N_w \cr
              B^{AB} & \text{ if } \deg_{sc} u^A u^B > N_w
           \end{cases} 
\end{equation*}
Recall that we have assumed ascending ordering of $\deg_{sc} u^A$, and
so matrix $B$ is lower triangular with respect to anti diagonal $\deg_{sc}
u^A u^B = N_w$. This proves that $\det B$ equals to product of those
constants. Since, by the choice of coordinates, we know that
$\begin{vmatrix}B^{AB}\end{vmatrix}$ is nondegenerate we conclude,
that all those constants are nonzero. 
\end{proof}

\begin{remark}\label{rem:OnCasimir}
 Notice, that $\N^1=w_2$ annihilates the pencil of finite
 dimensional brackets: $\overline{\pb{f}{\N^1}}_\lambda=0$ for any
 $f$.
\end{remark}

\begin{proposition} \label{prop:FixPointSurface}
 Let $\G^{A,a}\left(u,\N\right)=\overline{\pb{u^A}{\N^a}}_2$.
Consider a variety ${\mathcal M}_r$ defined by equations
$\G^{A,2}\left(u,\N\right)=0$ and suppose that matrix $\begin{vmatrix}
\frac{\partial \G^{A,2}}{\partial u^B} \end{vmatrix}$ is not
degenerate on $\M_r$. Then 
\begin{equation*}
  \G^{B,b} \left(u,\N \right)_{\vert \M_r} =0  \qquad \forall B, b\,.
\end{equation*}
\end{proposition}
\begin{proof}
 By the implicit function theorem one may resolve the system of
polynomial equations $\G^{A,2}\left(u,\N \right) =0$ locally with
respect to variables $u$. Consider the Hamiltonian flows in involution
generated by $\N^a$:
\begin{equation*}
 \frac{\partial \varphi}{\partial t_a} =
 \overline{\pb{\varphi}{\N^a}}_2 \,.
\end{equation*}
 Then one has
\begin{equation*}
  \frac{\partial}{\partial t_a} \frac{\partial}{\partial t_2} u^A =
  \frac{\partial}{\partial t_2} \frac{\partial}{\partial t_a} u^A
  \quad \Rightarrow \quad \frac{\partial \G^{A,2}}{\partial u^B}
  \G^{B,a} = \frac{\partial \G^{A,a}}{\partial u^B} \G^{B,2} \,.
\end{equation*}
 Restricting on $\M_r$ and using the non degeneracy of $\frac{\partial
\G^{A,2}}{\partial u^B}$ the result follows. 
\end{proof}

\begin{remark} \label{rem:FlowsFP}
 So, the equation $\G^{A,2}=0$, subject to condition of Proposition
 \ref{prop:FixPointSurface} defines an algebraic variety $\M_r$ of
 stationary points of all Hamiltonians in involution with respect to
 Poisson bracket \eqref{eq:FDbr}. $\M_r$ ``corresponds'' to the kernel
 of Poisson tensor of \eqref{eq:FDbr}.
\end{remark}

 As an immediate consequence of eqs.~\eqref{eq:Obv1} and
\eqref{eq:Obv2} we have the following
\begin{proposition} \label{prop:NoNr}
 Polynomials $\G^{A,a}\left(u,\N\right)$ do not depend on $\N^r$.
\end{proposition}

As follows from generalized Drinfeld-Sokolov construction, this finite
dimensional bracket is but Kirillov-Kostant bracket
\begin{equation}
  \overline{\pb{\varphi}{\psi}}_\lambda = \left( \Lambda+q \vert
  \left[ d_q \varphi \,, d_q \psi \right] \right) \label{eq:FDpencil}
\end{equation}
restricted on the space of $\ad P$ invariant functions $\varphi \left(\Tilde{q} \right) = \varphi \left(q \right)$:
\begin{equation}
  \Lambda + \Tilde{q} = \exp \left[ \ad n \right] \left( \Lambda +q
  \right) \qquad n \in P \label{eq:FDgaugeact}
\end{equation}

\begin{lemma} \label{lem:DSgen}
 Drinfeld-Sokolov coordinates $w^i(q)$, in dispersionless limit,
 generate the ring of $\ad P$ invariant polynomials in $q$.
\end{lemma}
\begin{proof}
Consider a ring of polynomial functions in $q$. It is obviously
generated by elements of $Q^\ast$. Using the adjoint action of
nilpotent group with Lie algebra being $P$ one may always reduce
$\Tilde{q}$ to canonical form $q \in Q^{can}$ (see
eq.~\eqref{eq:Qcan}). Indeed, projecting eq.~\eqref{eq:FDgaugeact} on
$\ad \rho$ eigenspaces we get
\begin{equation}
   \Tilde{q}_i = q_i + \left[  n_{i+1} , I_+ \right] +
   \text{polynomial}\left( n_i,\dots,n_1; q_{i-1},\dots,q_1 \right)
   \label{eq:FDQcanEq}
\end{equation}
where $\left[\rho,q_i\right]+i q_i =0$ and $0\leqslant i < N_w$. This
allows to resolve, starting from $i=0$ and proceeding inductively, for 
$\Tilde{q}$ and $n(q)$ as polynomials in $q$. Recapitulating,
$(\Tilde{q},n)$ generate a ring of polynomials in $q$. The subring of
gauge invariant polynomials is, then, obviously generated by
$\Tilde{q}$, i.e. by Drinfeld-Sokolov coordinates. 
\end{proof}

Now consider the case of modified hierarchy, $\s=\s_w$. The leading
term Poisson structure gives Kirillov-Kostant bracket on
$\ggg_0(\sw)^\ast$:
\begin{equation}
   \overline{\pb{\nu^i}{\nu^j}}_m = {f^{ij}}_k \nu^k\,. \label{eq:FDKiCoBr}
\end{equation}
The Miura map \eqref{eq:MiuraMap} provides a map from modified
hierarchy ($\s=\s_w$) into $\s=\s_h$ one. Thus we have polynomial
expressions $w^i\left(\nu\right)=w^i\left(q_m\right)$ for
dispersionless Drinfeld-Sokolov variables. Notice, that modified
hierarchies have their ``gauge algebra'' $P$ empty, and yet Miura
coordinates $\nu$ do not generate the ring of gauge invariant
polynomials. The reason is that, imposing $\Tilde{q} \in Q_m$,
equations \eqref{eq:FDQcanEq} do not have a unique solution.
Following ideas of ref.~\cite{BalFeh} we have

\begin{proposition}\label{prop:DiscreetR}
 There is a finite subgroup $R \subset \exp\left(\ad
 P\right)$ acting on $Q_m=\gg_0$.
\begin{equation}
   \ord R \leqslant \prod_{k \in \Pr_w} \left(k+1 \right)
   \label{eq:ordR} .
\end{equation}
\end{proposition}
\begin{proof}
 Let $n \in P$ corresponds to group element fixing $Q_m$, i.e.
\begin{equation}
  I_+ + \Check{q}_m = \exp\left[\ad n\right] \left(I_+ + q_m \right)
  =  \sum_{k=0}^{N_w+1} \frac{1}{k!} \ad^k n \left( I_+ + q_m \right).
  \label{eq:GaugeTrans}
\end{equation}
 Projecting on $\ad \rho$ eigenspaces we gain the system of
equations~\eqref{eq:FDQcanEq}. Since $q_m \in \gg_0$, we have
\begin{equation}
   \Check{q}_m = q_m + \left[ n_{1}, I_+ \right] \label{eq:Trans} .
\end{equation}
 The rest of equations~\eqref{eq:FDQcanEq} for $i>0$ determines
 $n$. Given positive $i$ there are $\dim \gg_{-i}$ scalar equations.
 Due to injectivity of map $\ad I_+ \colon \gg_{k} \to \gg_{k+1}$ for
 negative $k$ we may unambiguously solve for $\dim g_{-i-1}$ unknowns
 contained in $n_{i+1}$ and remain with $m_i=\dim \gg_{-i} - \dim
 g_{-i-1}$ scalar equations for $n_j$, where $j< i$.  Notice, that by
 definition, $m_i\not=0$ iff $i \in \Pr_w$ with $m_i$ being the
 multiplicity of $i$ in $\Pr_w$.  Starting with $i=1$ case
\begin{equation}
   \left[n_2 ,I_+ \right] + \left[ n_1, q_m + \frac{1}{2} \left[n_1,
   I_+ \right] \right] =0 \,, \label{eq:Forn2}
\end{equation}
 we solve for $n_2=n_2\left(n_1,q_m\right)$. Proceeding further with
 excluding $n_k$, $k>1$, we end up with $m_i$ equations of order
 $i+1$, for each distinct $i \in Pr_w$, to determine $\dim \M$
 unknowns contained in $n_1$. Notice, that the number of equations $\ord
 \Pr_w$ equals to the number of unknowns $\dim \gg_{-1}=\dim \M$ and
 by Bezout theorem we obtain no more than $\prod_{k \in \Pr_w} \left(
 k+1 \right)$ solutions $n_1=n_1\left(q_m\right)$, which determine
 that number of nonlinear transformations $q_m \to \Check{q}_m$ of
 $Q_m$. If only those $\dim \M$ equations for $n_1$ are independent we
 obtain an equality in eq.~\eqref{eq:ordR}. 
\end{proof}

Since $n_{1} \in g_{-1}$ completely determines any transformation from
$R$, as follows from the proof of Proposition~\ref{prop:DiscreetR}, it
is tempting to make an anz\"ats for the simplest transformations as $n
\in \gg_{-1}$. It proves to be consistent only for simply laced Lie
algebras $\gg$. We may then rewrite
eq.~\eqref{eq:GaugeTrans} as follows
\begin{equation*}
  \Check{q}_m = q_m+\left[n,I_+\right] + \sum_{k=1}^{N_w} \frac{1}{k!}
  \ad^k n \left( \frac{1}{k+1}\left[n,I_+\right] + q_m \right) .
\end{equation*}
To make the sum, contributing to negative $\ad \rho$ eigenspaces,
vanish we need that
\begin{subequations}\label{eq:FDSimplCond}
\begin{gather}
  \left[n,\left[ n , H \right]\right] = 0\,, \qquad \forall H \in
  \ggg_0\left(\sw\right) , \label{eq:ad2nVanish} \\ \left[n , q_m
  +\frac{1}{2}\left[n,I_+\right]\right] =0\,.  \label{eq:EqOnn}
\end{gather}
\end{subequations}

 To solve equations~\eqref{eq:FDSimplCond} we make use of Weyl group
$\Wgz$ of semisimple Lie subalgebra $\gg_0 \subset \gg$.
\begin{lemma}\label{lem:OrbitComm}
  Subspaces $\gg_{k} \subset \gg$, $k \in \Zbb$, are stable under
natural action of $\Wgz \subset \Wg$. Each orbit ${\mathcal Z} \subset
\gg_{-1}$ of $\Wgz$ is a \emph{commutative} subalgebra of $\gg$, if
$\gg$ is simply laced.
\end{lemma}
\begin{proof}
 The action of Weyl group $\Wgz$ are inner in $\gg_0$ and in $\gg$.
For any root $\alpha$, such that $E_{\alpha} \in \gg_0$, one has a
reflection $r_{\alpha}$ in the hyperplane perpendicular to the root
acting on $\gg$ as follows:
\begin{equation}
  \Hat{r}_\alpha = \exp\left[ \ad E_\alpha \right] \exp\left[ -
  \frac{2}{\alpha^2} \ad
  E_{-\alpha} \right]\exp\left[ \ad E_\alpha \right] \label{eq:refl1} .
\end{equation}
 These reflections act canonically on Cartan subalgebra $\hh \subseteq
 \gg_0 \subset \gg$. Since all elements of $\Wgz$ can be expressed as
 a product of these, we conclude that $\Wgz$ stabilizes $\gg_k$.  Take
 a root $\beta$, such that $E_\beta \in \gg_{-1}$, and consider the
 $\Wgz$ orbit ${\mathcal Z}$ that passes through it. Fix $w \in \Wgz$
 and assume that $w$ does not fix $\beta$. It suffices to prove that
 $\beta + w\left(\beta\right)$ is never a root.

 Since $\Wgz$ stabilizes $\gg_{-1}$, $\gamma \stackrel{{\rm def}}{=} w
 \left( \beta \right)- \beta$ is a linear combination of simple roots
 of $\gg_0$. For simply laced Lie algebras $\gg$, $\gamma$ itself is a
 root such that $E_\gamma \in \gg_0$. 
 Assume that $\beta + w\left(\beta\right) = 2 \beta + \gamma$ is a
 root, then
\begin{equation}
   \frac{2 \left( \beta \vert \gamma \right)}{\left( \beta \vert \beta
   \right)} \leqslant -2 \label{eq:GammaCond} \,.
\end{equation}
 This is impossible for any two roots $\beta$ and $\gamma$ of simply
 laced simple Lie algebra $\gg$. 
\end{proof}

\begin{corollary}\label{cor:g0toZ}
  For any orbit ${\mathcal Z}$, and any element $E_\beta \in {\mathcal
Z}$,  $\ad E_\beta$ maps $\gg_0$ on ${\mathcal Z}$.
\end{corollary}
\begin{proof}
  Consider the set ${\mathcal S} \stackrel{{\rm def}}{=} \left\{
  \left[X,Y\right] \, \vert \, X \in \gg_0 \,, Y \in {\mathcal Z}
  \right\}$. Since $\hh \subseteq \gg_0$ we have ${\mathcal Z}
  \subseteq {\mathcal S}$. The set ${\mathcal S}$ is stable under the
  homomorphism $\Wgz$ because $\gg_0$ and ${\mathcal Z}$ are stable.
  So ${\mathcal S}={\mathcal Z}$ and, hence, $\ad E_\beta$ maps
  $\gg_0$ in ${\mathcal Z}$. The surjectivity is obvious. 
\end{proof}
\begin{remark} \label{rem:Zroots}
 Having the simple roots chosen in $\hh^\ast$, it is plain to see,
 that different orbits ${\mathcal Z}$ may be labeled by simple roots
 of $\sw$ degree one, i.e. such $\alpha$ that $E_\alpha \in \gg_1$.
\end{remark}

 Choose an orbit ${\mathcal Z} \subset \gg_{-1}$, and let $\{X_q\}$
 denote the set of root vectors that form the basis in ${\mathcal Z}$.
 We then solve \eqref{eq:ad2nVanish} by letting $n=\sum x_q X_q \in
 {\mathcal Z}$. Due to Corollary \ref{cor:g0toZ} we have
 
\begin{lemma} \label{lem:RefPoint}
 Condition~\eqref{eq:EqOnn} gives $\dim {\mathcal Z}$ homogeneous
quadratic equations for parameters $x_i$, that posess non trivial solutions.
\end{lemma}

Let us speculate a bit on the structure of the group $R$. Fix $q_m \in
\gg_0$, some orbit ${\mathcal Z}$, and let $n\left(q_m\right)\in
{\mathcal Z}$ be a solution of eq.~\eqref{eq:EqOnn}, that gives the
map $n\colon q_m \to \Check{q}_m$. Apply to the result another
transformation with $\Check{n} \in {\mathcal Z}$. We get another map
$\Check{n} \colon \Check{q}_m \to \Check{\Check{q}}_m$. Notice that
their composition would be another map given by $n'\left(q_m\right) =
n\left(q_m\right) + \Check{n} \left( \Check{q}_m \left( q_m \right)
\right)$, $n' \colon q_m \to \Check{\Check{q}}_m$. Both $n$ and $n'$
will verify eqs.~\eqref{eq:FDSimplCond} and obviously $n \not\equiv
n'$. It is clear, that further iterations will again bring another
solution of \eqref{eq:EqOnn}. We therefore arrive to
\begin{proposition}
 For each $\Wgz$ orbit ${\mathcal Z} \subset \gg_{-1}$ there is subgroup
$R_{\mathcal Z}$ of $R$.
\end{proposition}
Notice, that if $\dim {\mathcal Z}=1$, $n'$ must vanish and our
simplest transformation must be reflections, i.e.  $R_{\mathcal Z} =
\Zbb_2$.

 Any transformation $n \in P$ from $R$ is uniquely determined by its
$\gg_{-1}$ projection $n_{-1}$. That $n_{-1}$ may be uniquely split
$n_{-1} =\sum_k n_{-1}^{(k)}$ with $n_{-1}^{(k)} \in {\mathcal Z}_k$
in distinct $\Wgz$ orbits. It suggests that
\begin{equation*}
   \exp\left[\ad n\left(q_m\right) \right]\left( I_+ +q_m \right) =
   \prod_i \exp \left[ \ad n_{-1}^{k_i}\left(q_m\right) \right]\left(
   I_+ +q_m \right) .
\end{equation*}
  To ensure that $R$ is generated by the simplest transformations one
must prove that each factor preserves $\gg_0$. We, however, do not have 
the proof. 

\begin{remark}
 In the case of standard Drinfeld-Sokolov hierarchy we have $\gg_0 =
\hh$. Then eq.~\eqref{eq:ad2nVanish} implies that $n = x_k
E_{-\alpha_k} \in \gg_{-1}$ and eq.~\eqref{eq:EqOnn} requires
\begin{equation}
   x_k = \frac{2}{\left( \alpha_k \vert \alpha_k \right)} \sum_{i=1}^r
   \left( \alpha_k \vert \alpha_i \right) \nu^i \quad \Rightarrow
   \quad r_k \left(\nu^j\right)= \nu^j - \frac{2 \delta_{jk}}{ \left(
   \alpha_k \vert \alpha_k \right)}\sum_{i=1}^r \left( \alpha_k \vert
   \alpha_i \right) \nu^i \,.  \label{eq:StandCox}
\end{equation}
 The same transformation of $\nu$ results from the Weyl reflection,
acting on $\gg_0 = \hh$, corresponding to simple root $\alpha_k$.
Hence, polynomials $w^i\left(\nu\right)$ result to be Coxeter
polynomials~\cite{BalFeh}.
\end{remark}

\begin{corollary}
  Polynomials $w^i\left(\nu\right)$ of degrees $k_i +1$ with $k_i \in
\Pr_w$ are invariant with respect to discrete group $R$ and generate
the ring of $R$ invariant polynomials on $Q_m$.
\end{corollary}
\begin{proof}
  By Lemma \ref{lem:DSgen} $w^k(q)$ generate the ring of gauge
invariant polynomials in $q$. Restricting them on $Q_m$ and
restricting adjoint group to $R$ the proof follows. 
\end{proof}

\begin{remark}
  Polynomials $w^i\left(\nu\right)$ are not however $\Wgz$
 invariant. This means that they are not, for non trivial $\Wgz$, a
 restriction to $\gg_0$ of ${\mathop {\rm Ad}}$ invariant polynomials
 on $\gg$.
\end{remark}

 Notice, $R$ invariance of $w^i\left(\nu\right)$ implies that
Kirillov-Kostant brackets is equivariant with respect to action of
$R$, i.e.
\begin{equation*}
   \overline{\pb{\Check{\nu}^i}{\Check{\nu}^j}}_m = {f^{ij}}_k
   \Check{\nu}^k = \frac{\partial \Check{\nu}^i}{\partial \nu^m}
   {f^{mn}}_l \nu^l \frac{\partial \Check{\nu}^j}{\partial \nu^n} \,.
\end{equation*}

 The $R$ invariant variety $\M_r$ is defined in $Q_m$, much the same,
as fixed point of all Hamiltonians in involution:
\begin{equation}
  F^k\left(\nu\right)= \overline{\pb{\nu^k}{\N^2}}_m = {f^{kl}}_n
   \nu^n \frac{\partial \N^2}{\partial \nu^l} \,. \label{eq:MrInMiuraCoord}
\end{equation}
 Indeed, \eqref{eq:MrInMiuraCoord} implies $\G^{A,2}=0$ and so defines
$\M_r$. It means, that one can restrict action of $R$ group on $\M_r$.

 To construct ``good'' coordinates on $\M_r$ we consider Casimirs of
Kirillov-Kostant bracket \eqref{eq:FDKiCoBr}. Recall, that $\gg_0$ is
semisimple Lie algebra, i.e $\gg_0 = \oplus_i {\gg'}^{(i)}$, where
${\gg'}^{(i)}$ is either simple or abelian.  Let $J_{i,k}(\nu)$ be
algebraically independent $\ad \gg_0$ invariant polynomials of degree
$k$ corresponding to ${\gg'}^{(i)}$, i.e. invariant also under $\ad
{\gg'}^{(i)}$. They obviously annihilate \eqref{eq:FDKiCoBr}.

Introduce $r$ ``abelian'' coordinates $\mu$ on $\hh^\ast = \oplus_i
\left({\hh}^{(i)}\right)^\ast$ defined by equations
\begin{equation}
   J_{i,k}\left(\mu\right)_{\vert \hh} =
  J_{i,k}\left(\nu \right) \,.  \label{eq:AbelCoord}
\end{equation}
 Coordinates $\mu$ have scaling weight one, as well as Miura
coordinates. 
\begin{proposition}
  Coordinates $\mu$ do not depend on the choice of $\ad \gg_0$
invariant polynomials $J_{i,k}$. 
\end{proposition}
\begin{proof}
 Indeed, let us choose another set of algebraically independent
polynomials 
\begin{equation*}
f_j\left(\nu\right) = f_j
\left(J\left(\nu\right)\right).
\end{equation*}
 They are known to be of the same degrees. Using them we define new
coordinates $\mu'$:
\begin{equation*}
   f_j\left(\mu'\right)\vert_{\hh^\ast} = f_j \left(\nu\right) \qquad
   f_j\left(J\left(\mu'\right)\right) \vert_{\hh^\ast} = f_j
   \left(J\left(\nu\right)\right) = f_j\left(J\left(\mu\right)\right)
   \vert_{\hh^\ast} \,.
\end{equation*}
 Starting from functions $f$ of the lowest degree and proceeding up we
 conclude that $\mu'$ may be chosen to coincide with $\mu$. 
\end{proof}
We may complete $\mu$ by those Miura coordinates $\nu$ that
correspond to $X_i \not\in \hh$ in $q_m = \sum \nu^i X_i$, to have
coordinates on the whole phase space $\M$. Let us denote them $\eta^i$
so that $\eta^i = \mu^i$ for $1 \leqslant i \leqslant r$.

\begin{lemma} \label{lem:Restr}
  Group $R$ admits restriction on the ``abelian'' coordinates $\mu$.
\end{lemma}
\begin{proof}
  Indeed, Casimirs $J_{i,k}$, being invariant with respect to adjoint
action of $\gg_0$, are not left invariant by action of group
$R$. Let $\Check{J}_{i,k} \left(\nu\right) = J_{i,k}\left(
\Check{\nu}\left(\nu\right) \right)$. Then
\begin{equation*}
\begin{split}
 0 &= {f^{lm}}_n \Check{\nu}^n \frac{\partial
 J_{i,k}\left(\Check{\nu}\right)}{\partial \Check{\nu}^m} =
 \overline{\pb{\Check{\nu}^l}{J_{i,k}\left(\Check{\nu}\right)}}_m
 \quad \Rightarrow \\ 0 &= {f^{lp}}_n \nu^n \frac{\partial
 \Check{\nu}^m}{\partial\nu^p} \frac{\partial
 J_{i,k}\left(\Check{\nu}\right)}{\partial \Check{\nu}^m} = {f^{lp}}_n
 \nu^n \frac{\partial \Check{J}_{i,k}\left(\nu\right)}{\partial \nu^m} \,.
\end{split}
\end{equation*}
This implies that $\Check{J}_{i,k}\left(\nu\right)$ results $\ad
\ggg_0(\sw)$ invariant and hence depends on $\mu$ only.

\end{proof}
\pagebreak[2]
 In the next section we shall make use of the following 
\begin{lemma} \label{lem:ItIsConst}
  Define the following matrix
\begin{equation}
    \left({\mathbb K}^{-1}\right)^{mn} = \frac{\partial
    \eta^{m}}{\partial \nu^i} \left( {\mathcal K}^{-1} \right)^{ij}
    \frac{\partial \eta^{n}}{\partial \nu^j} \label{eq:newKil} \,.
\end{equation}
 Then its submatrix $\left({\mathbb K}^{-1}\right)^{kl}$ with $1
 \leqslant k,l \leqslant r$ is constant non degenerate matrix.
\end{lemma}
\begin{proof}
 Since $J_{i,k}$ are $\ad \gg_0$ invariant polynomials and so is
Killing metrics, we may make use of Chevalley theorem
\begin{equation}
 \frac{\partial J_1\left(\nu\right)}{\partial \nu^i} \left({\mathcal
 K}^{-1}\right)^{ij} \frac{\partial J_2\left(\nu\right)}{\partial
 \nu^j} = F_{1,2}\left(J\left(\nu\right)\right) \label{eq:Chevalley}\,.
\end{equation}
stating that $F$ is a polynomial in $\ad \gg_0$ invariant
polynomials $J_{i,k}\left(\nu\right)$. Now using definition of coordinates
$\mu$ \eqref{eq:AbelCoord} and changing variables to $\eta$ we obtain
\begin{equation}
  F_{1,2}\left(J\left(\mu\right)\right) = \frac{\partial
  J_1\left(\eta\right)}{\partial \eta^i} \left({\mathbb
  K}^{-1}\right)^{ij} \frac{\partial J_2\left(\eta\right)}{\partial
  \eta^j} = \sum_{m,n=1}^r \frac{\partial J_1\left(\mu\right)}{\partial \mu^m}
  \left({\mathbb K}^{-1}\right)^{mn} \frac{\partial
  J_2\left(\mu\right)}{\partial \mu^n} \,. \label{eq:Aux3}
\end{equation}
 By another Chevalley theorem $J\left(\mu\right)$ are invariant with
 respect to Weyl group of $\gg_0$.  This means that $r \times r$
 submatrix ${\mathbb K}^{-1}$ is inverse of Killing metrics of rank
 $r$ semisimple Lie algebra $\gg_0$ restricted to Cartan
 subalgebra. The latter coincides with Killing form of algebra
 $\gg$. Thus it is non degenerate due to celebrated Cartan's criterion
 of semisimplicity.  
\end{proof}
\begin{corollary}
   $\left({\mathbb K}^{-1}\right)^{ij} = \left({\mathcal
  K}^{-1}\right)^{ij}$ for $1 \leqslant i,j \leqslant r$.
\end{corollary}
\pagebreak[2]
\begin{lemma} \label{lem:LinAction}
  Group $R$ acts linearly on coordinates $\mu$. 
\end{lemma}
\begin{proof}
Consider $\Check{J}\left(\mu\right)
=J\left(\Check{\mu}\left(\mu\right)\right)$. Rewrite
\eqref{eq:Chevalley} as follows
\begin{equation*}
\begin{split}
  F_{1,2}\left(\Check{J}\left({\mu}\right)\right) &=
  F_{1,2}\left(J\left(\Check{\mu}\right)\right) = \sum_{m,n=1}^r
  \frac{\partial J_1\left(\Check{\mu}\right)}{\partial \Check{\mu}^m}
  \left({\mathbb K}^{-1}\right)^{mn} \frac{\partial
  J_2\left(\Check{\mu}\right)}{\partial \Check{\mu}^n} \\ &=
  \sum_{m,n,k,l=1}^r \frac{\partial \Check{J}_1\left(\mu\right)}{\partial
  \mu^m} \frac{\partial {\mu}^n}{\partial \Check{\mu}^l}
  \left({\mathbb K}^{-1}\right)^{kl} \frac{\partial {\mu}^n}{\partial
  \Check{\mu}^l} \frac{\partial \Check{J}_2\left(\mu\right)}{\partial
  \mu^n}\,.
\end{split}
\end{equation*}
 Since functions $F_{1,2}\left(J\right)$ are not altered we have the
same pairing and due to Lemma \ref{lem:ItIsConst} $R$ action preserves
constant non degenerate submatrix of the matrix ${\mathbb K}^{-1}$. It
is only possible if $R$ acts linearly. 
\end{proof}

\begin{remark}
 Since the action of $R$ on $\mu$ is defined via action on polynomials
$J\left(\mu\right)$, it can be determined  only up to $\ad \gg_0$,
i.e. only as 
\begin{equation*}
  \Check{J}\left(\mu\right) = J\left( A \cdot \mu \right) 
\end{equation*}
for some constant matrix $A$, determined up to transformations
\begin{equation*}
 \mspace{50mu} A \to w_1 A w_2\,, \qquad \text{where} \quad w_1,w_2
 \in \Wgz .
\end{equation*}
\end{remark}

\begin{lemma}\label{lem:ProjOnMu}
  Non trivial transformations from $R_{\mathcal Z}$ correspond to the
same matrix $A_{\mathcal Z}$ modulo this equivalence. 
\end{lemma}
\begin{proof}
 Fix an $\Wgz$ orbit ${\mathcal Z}$ in $\gg_{-1}$. It suffices to
 show that $\Wgz$ permutes solutions of eq.~\eqref{eq:EqOnn}.

 Notice, that $\Wgz$ invariant polynomials $J$, corresponding to
 abelian constituents of $\gg_0$ are linear and may be chosen to be
 corresponding Miura variables as follows from \eqref{eq:StandCox}.
 Then, due to Lemma \ref{lem:LinAction} $R$ groups acts on them
 linearly in terms of $\mu$. Solving eqs.~\eqref{eq:AbelCoord} for
 $\mu\left(\nu\right)$ we get just $\ord \Wgz$ solutions as follows
 from Bezout theorem. They correspond to different solutions of
 \eqref{eq:EqOnn}. 
\end{proof}

\begin{corollary}
  Eqs.~\eqref{eq:EqOnn} have non more than $\ord \Wgz$ non trivial solutions.
\end{corollary}

\begin{corollary}
  Matrix $A_{\mathcal Z}$ may be chosen to be a reflection,
  i.e. $A_{\mathcal Z}^2=1$. 
\end{corollary}
\begin{proof}
 Since $R_{\mathcal Z}$ is a subgroup, there exist two transformations
 product of which is identity. Since they correspond to the same
 matrix $A_{\mathcal Z}$, the proof follows. 
\end{proof}

\begin{theorem}\label{thm:WeylMonodr}
 Let $\Wgz$ be Weyl group of $\gg_0$. Assume $\gg$ simply laced. Then
$R\vert_{\M_r} \rtimes \Wgz \simeq \Wg$.
\end{theorem}
\begin{proof}
  Weyl group $\Wgz$ preserves metrics ${\mathbb K}^{-1}$, and is
generated by reflections \eqref{eq:StandCox} corresponding to simple
roots of zero $\sw$ grade.  Group $R$, restricted to $\M_r$, also
preserves the metrics and is generated by reflections associated
with simple roots of $\sw$ grade 1, due to Lemmas \ref{lem:OrbitComm}
and \ref{lem:RefPoint}. We conclude that $R\vert_{\M_r} \rtimes
\Wgz$ is a finite group generated by $r$ transformations associated to
simple roots, that preserve ${\mathbb K}^{-1}$. Since it is an inverse
of the Killing form of algebra $\gg$ and $\Wgz$ acts canonically
\eqref{eq:StandCox}, it follows that $R\vert_{\M_r} \rtimes \Wgz$
is isomorphic to Weyl group $\Wg$. 
\end{proof}

\section{Dispersionless limit} \label{sec:zerophase}

As was proved by I.~Krichever~\cite{Krichever}, the dispersionless
limit, or zero phase Whitham averaging, of the standard
Drinfeld-Sokolov hierarchies provides solutions to WDVV equations.

Introducing dispersion parameter $\e$ the Poisson structure of 
hierarchies reads
\begin{equation}
\begin{split}
 \pb{w^i(x)}{w^j(y)} &= \sum_{k \geqslant 0} \e^{k-1}
  \pb{w^i(x)}{w^j(y)}^{(k)} \\ &= \frac{1}{\epsilon} A^{ij}(w)
  \delta(x-y) + g^{ij}(w) \delta'(x-y) \\ &+ \Gamma^{ij}_k(w)
  \left(w^k(x)\right)' \delta(x-y) + {\mathcal O}(\epsilon)\,.
\end{split}
 \label{eq:limit}
\end{equation}
\begin{remark}
For the standard Drinfeld-Sokolov hierarchies $A\equiv 0$ because
$\dim \M =r$ and hence annihilators of the first Poisson structure may
be chosen as coordinates on the whole ${\mathcal M}$. Note that due to
Corollary \ref{cor:IndepHam} we only have $r \leqslant \dim {\mathcal
M}$ independent annihilators, thus appearance of $A$ term should be,
generally, expected.
\end{remark}

Let us pick up $w^i=\left(\N^a,u^A\right)$ as coordinates on $\left(
Q^{can}\right)^\ast$ as in section \ref{sec:FinDimGeom}. We assume
that they are obtained by ultralocal change of variables from
Drinfeld-Sokolov variables, i.e. contain no derivative terms.

 Due to eq.~\eqref{eq:Obv2} hierarchy time flows of $\N^a$ have
polynomial $\e$ expansion, but dynamics of $u^A$ coordinates does not
enjoy this property
\begin{eqnarray}
 \frac{\partial \N^a}{\partial t_{b}} &=& \pb{\N^a}{H_b}_2 =
 \partial_x A^a(b) +\Oe  \,, \nonumber \\ 
 \frac{\partial u^A}{\partial
 t_{b}} &=& \pb{u^A}{H_b}_2 = \f{1}{\e}
 \G_\alpha^{A,b}\left(u,\N\right) + \partial_x A^A\left(b\right) + \Oe
 \,. \label{eq:nongrad}
\end{eqnarray}
 Hence brackets $\pb{\N^a}{\N^b}$ admit dispersionless limit, while
others do not. $G^{A,b}$'s come from $A$ term in~\eqref{eq:limit} and
thus are responsible for fast dynamics of $u$ coordinates.  If $u$
coordinates evolved so as to vanish $\G^{A,a}$ identically, we would
obtain a well defined dispersionless limit of the hierarchy. As we
have seen in section \ref{sec:FinDimGeom} it happens on the algebraic
subvariety $\M_r \subset \M$.

We thus supplement Dubrovin-Novikov
prescriptions~\cite{DubrovinNovikov} for the restriction of the
Poisson structure~\eqref{eq:limit} on the slow - modulated zero phase
solutions by requirement of additional restriction on $\M_r$.  Dirac
bracket provides restriction of the Hamiltonian structure on $\M$ to
$\M_r$. We review briefly, for reader's convenience, the construction
of Dirac bracket, referring to~\cite{DiracBr} for details.

  Given constraint equation $\G^{A,2}=0$ defining $\M_r$ and local
coordinates $\N^a$ there, we introduce new ones
\begin{equation*}
   \Breve{\N}^a\left(x\right) = \N^a\left(x\right) + \sum_{A} \int dy
   \tau_{A}^a \left(x,y \right) \G^{A,2} \left(w(y)\right) .
\end{equation*}
such that $\Breve{\N}^a \vert_{\M_r} = \N^a\vert_{\M_r}$ and with
$\tau$ subject to condition
\begin{equation*} 
 \pb{\N^a\left(x\right)}{\G^{A,2}\left(w(y)\right)}\vert_{\M_r} = 0 \,.
\end{equation*}
 Looking for solution of $\tau$ as formal $\e$ series $\tau =
\sum_{m\geqslant 0} \e^m \tau^{(m)}$, the equation above for $\tau$ amounts
to the following
\begin{equation}\label{eq:CondOnTau}
 \pb{\N^a(x)}{\G^{A,2}(y)}^{(k)}_{\vert_{\M_r}} + \sum_{m=0}^k \int dz
 {\tau^{(m)}}_B^a\left(x,z\right)
 \pb{\G^{B,2}(z)}{\G^{A,2}(y)}^{(k-m)}_{\vert_{\M_r}} =0 \,.
\end{equation}
 Due to Proposition \ref{prop:FixPointSurface} we obtain that
 $\tau^{(0)}=0$ provided that the matrix
\begin{equation*}
 \pb{\G^{B,2}(z)}{\G^{A,2}(y)}^{(0)}\vert_{\M_r}
\end{equation*}
is nondegenerate.

\begin{definition}
 Dirac bracket on $\M_r$ is defined as 
\begin{multline}
   \pb{\N^a(x)}{\N^b(y)}_D =
   \pb{\Breve{\N}^a(x)}{\Breve{\N}^b(y)}\vert_{\M_r} = \\ 
   \pb{\N^a(x)}{\N^b(y)}\vert_{\M_r} -  \int dz_1\, dz_2 \tau^a_A
   \left(x,z_1\right)
   \pb{\G^{A,2}\left(x\right)}{\G^{B,2}\left(z_2\right)}\vert_{\M_r}
   \tau^b_B \left(y,z_2\right) \label{eq:DiracBrDef}
\end{multline}
\end{definition}
 Dirac bracket verifies~\cite{DiracBr} Jacobi identity.  As an immediate
 consequence of \eqref{eq:DiracBrDef} we have the following
\begin{lemma}\label{lem:ift0vanish}
  If $\tau^{(0)}=0$ then
\begin{equation}
  \pb{\N^a(x)}{\N^b(y)}_D^{(k)} = \pb{\N^a(x)}{\N^b(y)}^{(k)}
  \vert_{\M_r} \, \qquad k=0,1\,. \label{eq:WhatWeNeed}
\end{equation}
\end{lemma}
Because of \eqref{eq:Obv2} the $\e \to 0$ expansion of Dirac bracket
\eqref{eq:DiracBrDef} starts with $k=1$ term, and thus the bracket
admits dispersionless limit. 
\begin{corollary}
 Dispersionless limit of bi Hamiltonian structure is bi Hamiltonian. 
\end{corollary}
We thus arrive to the following theorem
\begin{theorem} \label{thm:Maltsev1}
 Consider Hamiltonian dynamical system admitting constant solutions.
Let some $\N^a$ be the densities of the local commuting integrals of
the system, considered as the parameters of the full family of the
constant solutions.  Let $u$ denote the rest of dynamical variables.
Assume that the matrix $\overline{\pb{\G^{A,2}}{\G^{B,2}}}_{2}
\vert_{\M_r}$ does not degenerate identically. Then the dispersionless
limit of the Hamiltonian structure restricted to $\M_r$, given by
Dubrovin-Novikov formula
\begin{equation}
 \pb{\N^a(x)}{\N^b(y)}^\ast = g^{ab}\left(\N(x)\right) \delta'(x-y) + 
  \Gamma^{ab}_c \left( \N(x) \right) \left(\N^c(x)\right)' \, \delta(x-y) \,,
 \label{eq:hydrobr}
\end{equation}
satisfies the Jacobi identity and does not depend on the choice of $\G$.
\end{theorem}
 This result is a particular case of theorem due to A.~Maltsev
\cite{Maltsev} who proved, using Dirac reduction procedure, that
Dubrovin-Novikov averaging procedure yields, under certain
assumptions, a Poisson structure on the space of $m$-phased solutions
of dynamical equations of original Hamiltonian system.

\pagebreak[2]
\begin{lemma}
  Assumptions of Theorem \ref{thm:Maltsev1} verify for the hierarchies
$(\gg,[w],\Lambda)$ with regular primitive conjugacy class $[w]$ and
grade one regular element $\Lambda$.
\end{lemma}
\begin{proof}
 Due to \eqref{eq:Viral}, $H_\Lambda=\int dx w_2\left(x\right)$ is the
momentum for the hierarchies in question, so their dynamical equations
admit constant solutions, because of theorem~\ref{thm:commofflows}.

Now we address the non degeneracy statement.
\begin{equation*}
{\mathcal A}^{AB}=\overline{\pb{\G^{A,2}}{\G^{B,2}}}_{2}
 =  \frac{\partial\G^{A,2}}{\partial w^i} \overline{\pb{w^i}{w^j}}_2 
\frac{\partial\G^{B,2}}{\partial w^j} \,.
\end{equation*}
 Restricting on $\M_r$ and using \eqref{eq:Obv2} and 
 $\overline{\pb{u^A}{\N^a}}_2\vert_{\M_r}=0$ we conclude 
\begin{equation*}
  {\mathcal A}^{AB}\vert_{\M_r} = \frac{\partial\G^{A,2}}{\partial u^C}
\left( u\left(\N\right), \N \right) \overline{\pb{u^C}{u^D}}_2 \vert_{\M_r}
\frac{\partial\G^{B,2}}{\partial u^D}\left( u\left(\N\right), \N
\right) \,. 
\end{equation*}
 Thus, due to assumption of Proposition \ref{prop:FixPointSurface},
 it suffices to prove the non degeneracy of ${\mathbb
 A}^{CD}=\overline{\pb{u^C}{u^D}}_2 \vert_{\M_r}$. The latter is
 obvious because
\begin{equation*}
  {\mathbb A}^{CD} = \Tilde{{\mathbb A}}^{CD} + \N^r
  B^{CD}\vert_{\M_r} \,,
\end{equation*}
where $\Tilde{{\mathbb A}}$ does not depend on $\N^r$.  Restriction on
$\M_r$ does not alter the linear dependence of ${\mathbb A}$ on $\N^r$
because of Proposition \ref{prop:NoNr}. Due to Lemma
\ref{lem:detBconstant} $\det B \not=0 $ and is constant. It, thus,
remains constant after restriction to $\M_r$.  
\end{proof}

\begin{proposition}[\cite{DubrovinNovikov,DuFlat}]
Given Poisson structure of hydrodynamic type \eqref{eq:hydrobr},
$g^{ab}$ is the flat covariant metrics as long as $\det g \not=0$ and
$\Gamma$ is its connection, related to Levi-Civita connection by the
following relation
\begin{equation*}
 \Gamma^{ab}_c = - g^{ad} \Gamma_{dc}^b \,.
\end{equation*}
\end{proposition}

 Non degeneracy of matrix $g^{ab}\left(\N\right)$ obtained by Dirac
reduction on $\M_r$ of Poisson structure of generalized integrable
hierarchies in consideration is not an obvious fact and needs to be
proved. 

 Miura map \eqref{eq:MiuraMap} provides us with polynomial expressions
 of Drinfeld-Sokolov coordinates in terms of Miura ones and its
 derivatives:
\begin{equation*}
  w^i_{\e}\left(\nu\right) = w^i\left(\nu\right) + \e
  \sum_{j=1}^{\dim \M} \left(\nu^j\right)' w^{(1)}_j \left(\nu\right) +
  {\mathcal O}\left(\e^2\right) \,.
\end{equation*}
 This change of coordinates provides a map of Miura bracket
 \eqref{eq:i1MiuPB} to the second Poisson structure
 \eqref{eq:i1BhSt}. Under this map we obtain for the first two matrix
 in \eqref{eq:limit} the following expressions
\begin{eqnarray}
  A^{ij}\left(w\right) &=& \frac{\partial w^i}{\partial \nu^k}
  {f^{kl}}_m \nu^m \frac{\partial w^j}{\partial \nu^l}\,, \nonumber \\
  g^{ij}\left(w\right) &=& \frac{\partial w^i}{\partial \nu^k}
  \left({\mathcal K}^{-1}\right)^{kl} \frac{\partial w^j}{\partial
  \nu^l} + \left( w^{(1)}_j \frac{\partial w^i}{\partial \nu^k} -
  w^{(1)}_i \frac{\partial w^j}{\partial \nu^l} \right) {f^{kl}}_m
  \nu^m \,. \label{eq:ImpOrtAnt}
\end{eqnarray}
 
As an immediate consequence of definition of $\M_r$ we have
\begin{proposition}\label{prop:Simpl}
 On $\M_r$ \eqref{eq:ImpOrtAnt} simplifies to the following
 expression:
\begin{equation}
 g^{ab}\left(\N\right) = \sum_{k,l=1}^{\dim \M} \frac{\partial
  \N^a}{\partial \nu^k} \left({\mathcal K}^{-1}\right)^{kl}
  \frac{\partial \N^b}{\partial \nu^l} \label{eq:Simpl} \,.
\end{equation}
\end{proposition}

\begin{proposition}\label{prop:Trivial}
 $\pb{\cdot}{\cdot}_1^\ast$ can be read off $\pb{\cdot}{\cdot}_2^\ast$
by appropriate shift of $\N_r$. 
\end{proposition}
\begin{proof}
 Due to Proposition \ref{prop:NoNr} and Lemma \ref{lem:ift0vanish}
dispersionless limit of the second bracket remains linear in $\N^r$.
The result follows.  
\end{proof}
 Note, that $\N_1=w_2$ satisfies Fuchs algebra
\begin{equation*}
 \pb{\N^1}{\N^1}_2^\ast = 2 \N^1 \delta' + \left(\N^1\right)' \delta
 \,.
\end{equation*}
Thus we have arrived to
\begin{proposition} \label{prop:BiHamPres}
  Zero phase Whitham averaging maps graded bi Hamiltonian
structure~\eqref{eq:i1BhSt} into graded bi Hamiltonian structure of
hydrodynamic type.
\end{proposition}
 Due to Lemma \ref{lem:ift0vanish} $\N^a$ will remain annihilators of 
the first bracket: 
\begin{equation}
 \label{prop:1stBrInCasimirs}
\pb{\N^a(x)}{\N^b(y)}_1^\ast = \eta^{ab} \delta' \left(x-y\right) .
\end{equation}

\begin{remark}
 Due to scaling weight grading, and chosen field ordering $\eta^{ab}$
 is anti-diagonal matrix.
\end{remark}
Indeed, $\eta^{ab}$ vanishes if $\deg_{sc} \left( \N^a \N^b \right)
\not= N_w +2$ and $\deg_{sc} \left( \N^a \N^{r+1-a} \right) = N_w+2$
due to $\deg_{sc} \N^a = k_a+1$ and ascending ordering of $k_a \in
I(w)$.
 Notice, that this is consistent with previously assigned
$\N^1$ being $w_2$ and $\N^r$ linear in $w_{N_w,1}$, because of
\begin{equation}
 \pb{\N^k}{\N^1}_2^\ast = \deg_{sc} \left( \N^k \right) \N^k \delta' +
                          \left(\N^k\right)' \,\delta \,.
                          \label{eq:Wdisp}
\end{equation} 

\begin{proposition}
 Dispersionless limit of Dirac restriction of Poisson structure of
 modified hierarchy reads
\begin{equation}
  \pb{\mu^i}{\mu^j}^\ast = \left({\mathbb K}^{-1}\right)^{ij} \delta'
  \left(x-y\right) \label{eq:FlatCoordPB} .
\end{equation}
\end{proposition}
\begin{proof}
 $\mu$ are Casimirs of finite dimensional Kirillov-Kostant bracket
 \eqref{eq:FDKiCoBr} and thus due to Lemma \ref{lem:ift0vanish} and
 Lemma \ref{lem:ItIsConst} the result follows.
\end{proof}

\begin{theorem}\label{thm:Main1}
  Metrics $g^{ab}\left(\N\right)$ is not identically
  degenerate. Coordinates $\mu$ are flat coordinates for this metrics.
\end{theorem}
\begin{proof}
 Let us choose coordinates $\eta$ as follows. Take the first $r$
coordinates to be $\mu$ and choose the rest to be canonical for finite
dimensional bracket. The advantage of this choice is the simplicity of
constraint equations \eqref{eq:MrInMiuraCoord}
\begin{equation}
    \left. \frac{\partial \N^2}{\partial \eta^k}\right|_{\M_r} =0
    \quad \overset{\text{ by Prop.
    \ref{prop:FixPointSurface}}}{\Rightarrow} \quad
    \left. \frac{\partial \N^a}{\partial \eta^k}\right|_{\M_r} =0
    \qquad \forall \, r+1 \leqslant k \leqslant \dim \M \,.
\end{equation}
Due to Lemma \ref{lem:ift0vanish} and Proposition \ref{prop:Simpl} 
we have 
\begin{equation*}
\begin{split}
  g^{ab} &= \left. \sum_{k,l=1}^{\dim \M} \frac{\partial
  \N^a}{\partial \nu^k} \left({\mathcal K}^{-1}\right)^{kl}
  \frac{\partial \N^b}{\partial \nu^l}\right|_{\M_r} 
  = \left. \sum_{k,l=1}^{\dim \M} \frac{\partial
  \N^a}{\partial \eta^k} \left({\mathbb K}^{-1}\right)^{kl}
  \frac{\partial \N^b}{\partial \eta^l}\right|_{\M_r} \\
  &= \left. \sum_{k,l=1}^{r} \frac{\partial
  \N^a}{\partial \mu^k} \left({\mathbb K}^{-1}\right)^{kl}
  \frac{\partial \N^b}{\partial \mu^l}\right|_{\M_r} \,.
\end{split}
\end{equation*}
 Both set of coordinates $\N$ and $\mu$ are local coordinates on $\M_r$,
 and so $J(\mu)= \det\begin{vmatrix} \frac{\partial \N^a}{\partial
 \mu^i}\end{vmatrix}$ does not degenerate at generic point of $\M_r$.
 This proves non degeneracy. 

 The following identity along with Lemma \ref{lem:ItIsConst} prove
 that $\mu$ are indeed flat coordinates of metrics $g$:
\begin{equation*}
   \frac{\partial \N^a\vert_{\M_r}}{\partial \mu^k} = \frac{\partial
   \N^a\left(\mu,\eta\left(\mu\right)\right)}{\partial \mu^k}=
   \left.\frac{\partial \N^a}{\partial \mu^k}\right|_{\M_r} +
   \sum_{i=r+1}^{\dim \M} \left.\frac{\partial \N^a}{\partial
   \eta^i}\right|_{\M_r} \left. \frac{\partial \eta^i}{\partial
   \mu^k}\right|_{\M_r} = \left.\frac{\partial \N^a}{\partial
   \mu^k}\right|_{\M_r} \,.
\end{equation*}
 As a byproduct we conclude that brackets \eqref{eq:hydrobr} and
\eqref{eq:FlatCoordPB} define the same geometry on $\M_r$:
\begin{equation}
   \pb{\N^a\left(x\right)}{\N^b\left(y\right)}^\ast_2 = \frac{\partial
   \N^a}{\partial \mu^m} \left(x\right)
   \pb{\mu^m\left(x\right)}{\mu^n\left(y\right)}^\ast_m \frac{\partial
   \N^b}{\partial \mu^n} \left(y\right) . \label{eq:TheSameGeom} \quad
   \text{on } \M_r
\end{equation}
 This conclusion may be drawn also noting that both coordinates set are
densities of mutually commuting integrals of corresponding exact
brackets, the property that survives averaging. 
\end{proof}

\begin{corollary}\label{cor:Inv}
 $\N^a\vert_{\M_r}\left(\mu\right)$ are invariant with respect to
 linear action of $R$ group on $\M_r$. 
\end{corollary}
\begin{proof}
 Indeed, due to Lemma \ref{lem:Restr} $R$ admits a restriction of
 $\M_r$ and preserves the latter. Hence, $\N^a$ were $R$ invariant in
 $\M$ and so they remain restricted on $\M_r$. 
\end{proof}
 
 Following K.~Saito~\cite{Saito} and using Theorem \ref{thm:Main1} we
 obtain the following
\begin{proposition}
  Metrics $\eta^{ab}$ is non degenerate.
\end{proposition}
\begin{proof}
 Consider the following polynomial in $\lambda$
\begin{equation}
\begin{split}
  P\left(\lambda\right) &= \det \left\vert g^{ab}\left(\N\right) -
  \lambda \eta^{ab} \right\vert = \det \left\vert g^{ab} \left( \N^1,
  \dots, \N^{r-1}, N^r - \lambda \right)\right| \\ &= \det
  \begin{vmatrix} \eta^{ab} \end{vmatrix} \left( \N^r - \lambda
  \right)^n+ \sum_{n=0}^{r-1} c_n\left(\N^1,\dots,\N^{r-1} \right)
  \left( \N^r - \lambda \right)^n \,.  \label{eq:Pl}
\end{split}
\end{equation}
When all $\N$ but $\N^r$ vanish it simplifies to $P(\lambda) = \det
\begin{vmatrix} \eta^{ab} \end{vmatrix} \left( \N^r - \lambda
\right)^n$. At this point $J\left(\mu\right) \not=0$ and thus $g^{ab}$
is non degenerate. Indeed, let $\mu$  be eigenvector of some
representative, which always exists, of $[w]$ in $R$ with eigenvalue 
$\xi=\exp\left[2 i \pi/N\right]$. Then, due to homogeneity of 
$\N^a(\mu)$ and from their
$R$ invariance, we conclude that only variables of degree $N$ may
differ from zero at this point of $\M_r$. If there are more than one
such variable, then eigenvalue $\xi$ is degenerate and we can always
choose $\mu$ so that only $\N^r$ does not vanish. Due to regularity
of $[w]$, vector $\mu$ is not left fixed by any transformation from
$R$, and thus the results follows. 
\end{proof}

\begin{proposition} \label{prop:quasihom}
 So obtained pencil of Hamiltonian structures provides us with 
quasihomogeneous~\cite{DuFlat} flat pencil of metrics. 
\end{proposition}
\begin{proof}
 Let $g$ be the metrics of the second Poisson structure and let $\eta$
-- of the first. Introduce function $\tau =\N^1 / N_w$, and
introduce the following vector fields
\begin{equation}
   E^a = g^{ab} \partial_b \tau \qquad e^a = \eta^{ab} \partial_b \tau
 \label{eq:vectfields} \,.
\end{equation} 
Notice, that with this choice $e^a = \delta_{a,r}$ and ${\mathfrak
L}_e g = \eta$ and ${\mathfrak L}_e \eta =0$ as follows from
considerations above, and where we have assumed $\eta$ to be chosen
anti diagonal with all nonzero entries being $N_w$. Then $E^{a} =
\deg_{sc} \left(\N^a\right) / N_w \N^a$, as follows
from~\eqref{eq:Wdisp}. One sees immediately that $\left[
e,E\right]=e$.  The second Poisson structure is scaling weight graded
and $N_w E$ is scaling weight Euler vector field. Thus $g$ must be
an eigenvector of ${\mathfrak L}_E$ : ${\mathfrak L}_E g = (d-1) g$.
In ref.~\cite{DuFlat} such flat pencils were called quasihomogeneous of
degree $d$.  
\end{proof}

\begin{theorem} \label{thm:solWDVV}
 Given $\pb{\cdot}{\cdot}_\lambda^\ast$ one may associate to it a
 solution to WDVV.
\end{theorem}
\begin{proof}
 Since obtained pencil of Hamiltonian structures satisfies Jacobi
identity we have a flat pencil of metrics. It is quasihomogeneous as
was shown in proposition~\ref{prop:quasihom}. Thus, following
ref.~\cite{DuFlat}, it is enough to show that the degree of
quasihomogeneity $d \not= 1$.

As was said in the proof of proposition~\ref{prop:quasihom} $E^a =
\tfrac{\deg_{sc} \N^a}{N_w} N^a$. Due to quasihomogeneity of flat pencil
we have ${\mathfrak L}_E \eta = (d-2) \eta$. But
\begin{equation*}
\begin{split}
 {\mathfrak L}_E \eta^{ab} &= E^c  \partial_c \eta^{ab} - 
                             \eta^{ac} \partial_c E^b - \eta^{cb}
                             \partial_c E^a \\
                           &= -  \left( \frac{ \deg_{sc}
                             \N^b}{N_w} + \frac{ \deg_{sc}
                             \N^a}{N_{w}} \right) \eta^{ab}  \\
                           &= - \frac{ N_w +2}{N_{w}} \eta^{ab} \,.
\end{split}
\end{equation*}
In the last line we have used the fact that $\eta^{ab}$ vanishes
unless $\deg_{sc} \left( \N^a \N^b \right) = N_{w}+2$, as follows
from scaling weight grading of Poisson structures. From this we obtain
\begin{equation*}
 d = 1 - \frac{2}{N_{w}} \,.
\end{equation*}
Since $N_w$ is finite we obtain that $d<1$. Note that for the
Coxeter conjugacy class $N_w = h$ - Coxeter number, and we recover
the formula of B.~Dubrovin, obtained while constructing polynomial
solutions to WDVV equations on the orbits of Coxeter
groups~\cite{DuCox}.

 Practically, we can find Frobenius potential $F(\N)$ from the following
relations
\begin{equation}
\begin{split}
  g^{ab}\left(\N \right) &= \left( d-1-d_a -d_b \right) \eta^{ac}
 \eta^{bd} \partial_c \partial_d F\left(\N\right) , \\
  \Gamma^{ab}_c  &= \left(
 \frac{3-d}{2}-d_a \right) \eta^{ad}\eta^{bf}\partial_d \partial_f
 \partial_c F \,,
\end{split}
\label{eq:toextract}
\end{equation}
where $d_a \dl_a^b = \partial_a E^b$. 
\end{proof}

 Thus, $\N^a$ are Saito coordinates~\cite{Saito} on Frobenius
manifold~\cite{DuRev} being flat coordinates for the metric $\eta$.
However, flat coordinates $\mu$ of the intersection metrics $g$ are also
very important. They clarify the geometric origin of the Frobenius
structure.  In general it is a challenging task, given a flat metric,
to find its flat coordinates. But in the case in question we were lucky
to use the theory of integrable systems.

\section{Example: $[w]=D_4(a_1)$}\label{sec:example}

To illustrate the developed technique we consider the example served
as the motivation of the present work. Let $\gg=D_4$ -- the simplest classical
Lie algebra where non Coxeter primitive conjugacy class
occurs (see appendix \ref{app:uno}). Luckily it enjoys regularity property.

 Take $[w]=D_4(a_1)$. We have readily that $I(w)=(1,1,3,3)$ and the
set of conformal weights $\Pr_w=(1,1,1,2,3,3)$. Miura coordinates 
form a Kac Moody algebra of $\ggg_0(\sw)= u(1)^{\otimes 3} \oplus
su(2)$, thus we shall have three exact Casimirs and one will be
computed in dispersion parameter expansion.

 Positive roots of $D_4$ read 
\begin{equation*}
\begin{split}
{\mathcal R}_+ &= \left\{ \alpha_1, \alpha_2, \alpha_3, \alpha_4,
 \alpha_1+\alpha_2, \alpha_2+\alpha_3, \alpha_2+\alpha_4,
 \alpha_1+\alpha_2+\alpha_3, \right. \\ 
 &{\mspace{8mu}} \left. \alpha_1+\alpha_2+\alpha_4, 
 \alpha_2+\alpha_3+\alpha_3, \alpha_1+\alpha_2+\alpha_3+\alpha_4, 
 \alpha_1+2 \alpha_2+\alpha_3+\alpha_4 \right\} .
\end{split}
\end{equation*}
Let us denote Lie algebra elements $E_\alpha$, $\alpha \in {\mathcal
R}_+$ by its decomposition on simple roots. So if
$\alpha=\alpha_{max}$ we write $X_{12234}$, and for
$E_{\alpha_1+\alpha_2}$ write $X_{12}$.  Similarly for negative roots,
substituting $X$ with $Y$. We fix Cartan-Weyl
basis~\eqref{eq:innp1}. So 
\begin{equation*}
\rho = \sum_{i,j=1}^r K^{-1}_{ij} s_j
H_{\alpha_i} = 2 H_{\alpha_1} + 3 H_{\alpha_2} + 2H_{\alpha_3}+ 2 H_{\alpha_4}  \,.
\end{equation*}
The Heisenberg subalgebra $\Hw$ is spanned by $z^k \Lambda_{i,1}$,
$z^k \Lambda_{i,2}$ for $i \in \{1,3\}$ and $k \in \Zbb$.
\begin{equation}
\begin{split}
 \Lambda_{1,1} &= X_1 + X_3 + z Y_{12234} + X_{12} + X_{23} + X_{24} \,,
 \\
 \Lambda_{2,1} &= X_1 - X_3 + X_4 - z Y_{1234} - X_{12} + X_{23} \,, \\
 \Lambda_{3,1} &= X_{1234} -\frac{z}{2} \left( Y_1 - Y_3 + 2 Y_4 -
 Y_{12} + Y_{23} \right) ,\\
 \Lambda_{3,2} &= -X_{12234} - \frac{z}{2} \left( Y_1 + Y_3 + Y_{12} +
 Y_{23} + 2 Y_{24} \right) .
\end{split}
\label{eq:Hwspan}
\end{equation}
These basis was just guessed, verifying linear independence and
regularity. It tur\-ned out easier than proceed as in
\eqref{eq:HeisenbergDef}.

Let us choose $\Lambda=\Lambda_{1,1}$ for our integrable
hierarchy. The $sl_2$ subalgebra constituents read $\rho, I_+, I_-$,
where $I_- = 3 Y_1+3 Y_3+Y_{12}+Y_{23}+4 Y_{24}$ and
$I_+=P_0^{\s_h}\Lambda$.
\paragraph{The hierarchy}
Let us fix the minimal weight gauge~\eqref{eq:MinWeightGauge} as follows
\begin{equation*}
\begin{split}
  q^{can} &= \left(w_2 - \frac{1}{10} u_2 - \frac{1}{2} v_2 \right)
   \frac{1}{12} I_- + u_2 \frac{1}{20} \left( 3 Y_1 - 3 Y_3 +
   12 Y_4 \right) + \\ &+ v_2 \frac{1}{20} \left(3 Y_1 + 3 Y_3 + 4Y_{24}
   \right)+ w_3 \frac{1}{6} \left( Y_{123} - Y_{234} - 2Y_{124}\right) + \\ 
    &+ Y_{1234} u_4 + Y_{12234} w_4 \,.
\end{split}
\end{equation*}
Hamiltonians, annihilators of the first Poisson structure, read
\begin{equation}
\begin{split}
 H_{\Lambda_{1,1}} &= \int dx \, w_2 \,, \qquad H_{\Lambda_{1,2}} = \int
 dx \, u_2 \,, \\
 H_{\Lambda_{3,1}} &= \int dx \, \left(-u_4+\frac {37}{3} u_2^2-\frac{1}{6}
 w_2 u_2+\frac{1}{64} w_2^2+\frac{1}{6} w_2 v_2- \right. \\ 
  &{} \mspace{200mu} - \left. \frac{7}{6}u_2 v_2+
 \frac{7}{12} v_2^2 \right) , \\
 H_{\Lambda_{3,2}} &= \int dx \, \left( w_4
 -\frac{7}{3}u_2^2+\frac{7}{24} w_2u_2+\frac{7}{3}u_2 v_2 \right) .
\end{split}
\label{eq:HamAnn}
\end{equation}
 We shall denote the densities of this annihilators as
 $\N_1$, $\N_2$, $\N_3$ and $\N_4$ respectively. 

As was explained in section~\ref{sec:typeI} fixing the gauge somehow,
we are able to compute both Poisson structures exactly, but the output
is enormous to be presented here. On the other hand we need then to
pass to $\N,w_3,v_2$ coordinates and eliminate auxiliary coordinates
$w_3$ and $v_2$. To do so we need to know $\Ww$ algebra at least up to
${\mathcal O}\left(\e\right)$ after rescaling. But we choose to we
omit these intermediate steps due to space restrictions and present the answer.
 
 The simplest Hamiltonian generating $G$ terms is $H_{\Lambda_{1,2}}$. Its 
flows with respect to the second Hamiltonian structure read
\begin{equation}
\begin{split}
 \f{\partial \N_1}{\partial t_{\Lambda_{1,2}}} &= \N'_2 \,, \qquad \quad 
 \f{\partial \N_2}{\partial t_{\Lambda_{1,2}}} = \f{1}{3} \left( v_2 -\N_2 + 5
 \N_1 \right)' \,, \\
 \f{\partial \N_3}{\partial t_{\Lambda_{1,2}}} &= 
 \f{1}{3} \left( \N_1 \N_2 + 3 \N_4 - \f{1}{4} \N_2^2 + \f{1}{4}v_2
 \N_2 + \e \f{2}{5} w'_3 \right)'\,, \\ 
 \f{\partial \N_4}{\partial t_{\Lambda_{1,2}}} &= 
   \f{1}{3} \left( \N_3 + \f{1}{4} \N_2^2 +
 \f{1}{12} \N_1\left(\N_1 -\N_2 +v_2\right) \right)' \,, \\ 
 \f{\partial v_2}{\partial t_{\Lambda_{1,2}}} &= \f{12}{\e} w_3 +
 \f{1}{3} \left( 5 \N_1 + 22 \N_2 + 2 v_2 \right)'\,, \\
 \f{\partial w_3}{\partial t_{\Lambda_{1,2}}} &= \f{1}{ \e } 
 \left( - 4 \N_3 + \f{1}{12} v_2^2 +\f{1}{3}  \N_1 v_2 - 
 \f{1}{3} \N_1 \N_2 - \f{1}{6} \N_2 v_2 - \f{1}{6} \N_2^2 +
 \f{1}{4} \N_1^2 \right) + \\ &+ \f{3}{20} \e \left( \N_2 - v_2 \right)'' \,.
\end{split}
\label{eq:N2flows}
\end{equation}
 Following the recipe, we take $1/\e$ terms as constraints. Using them
we obtain equation for the phase space subvariety $\M_r$ of slow motion
\begin{equation}
  w_3 =0\,, \qquad v_2 = \N_2 - 2 \N_1 \pm \Del \,, \qquad \Del = \sqrt{
  \N_1^2 + 3 \N_2^2 + 48 \N_3} \,.
\label{eq:uofN}
\end{equation}
This finally leads to the following restricted bi Hamiltonian
structure
\begin{equation}
\begin{split}
  \pb{\N_i(x)}{\N_j(y)}_1^\ast &= 4 \dl_{i+j,5} \dl'\left(x-y\right)
  \,.  \\
   \pb{\N_i(x)}{\N_j(y)}_2^\ast &= \left[ \gamma^{ij}\left(\N(x)\right) +
   \gamma^{ji}\left(\N(y)\right) \right]  \dl'\left(x-y\right) ,
\end{split}
\label{eq:solut}
\end{equation}
 where matrix $\gamma(\N)$ reads
\begin{gather*}
     \gamma^{i,1}=\{ \N_1, \N_2, 3 \N_3, 3\N_4 \} \,,\qquad
     \gamma^{1,i}=\{ \N_1, \N_2, \N_3, \N_4 \} \,, \\
  \gamma^{2,2} = \frac{1}{3} \left( \N_1 + 2 \Del \right) , \qquad 
  \gamma^{2,3} =  \N_4 + \f{1}{6} \N_1 \N_2 + \f{1}{12} \N_2 \Del \,,
     \qquad \gamma^{3,2} = 3 \gamma^{2,3} \,, \\
  \gamma^{2,4} =  \f{1}{3} \left( \N_3 - \f{1}{12} \N_1^2 + \f{1}{4}
     \N_2^2 + \f{1}{12} \N_1 \Del \right) , \qquad \gamma^{4,2} = 3
     \gamma^{2,4} \,, \\
  \gamma^{3,3} = \f{1}{2} \left( \N_1 \N_3 + \f{3}{32} \N_1 \N_2^2 +
     \f{7}{288} \N_1^3 \right) + \f{1}{288} \left( \N_1^2 +12 \N_2^2
     +48 \N_3 \right) \Del \,, \\
   \gamma^{4,3} = \f{1}{2} \left( \N_2\N_3 + \f{7}{96} \N_2\N_1^2
     +\f{1}{32}\N_2^3 +\f{1}{48} \N_1\N_2\Del \right) , \qquad
     \gamma^{3,4}=\gamma^{4,3} \,, \\
   \gamma^{4,4} = \f{1}{6} \left( -\N_1\N_3+\f{19}{288} \N_1^3+
     \f{7}{32}\N_1\N_2^2 + \f{1}{144} \left( 4\N_1^2 +3\N_2^2 +48\N_3
     \right)\Del\right) .
\end{gather*}
 The metric $g=\gamma+\gamma^{tr}$ is invertible,can be checked to be
flat and forms, obviously, together with $\eta$ a flat pencil.

\paragraph{Modified hierarchy}
 We consider the modified hierarchy to exemplify discrete group $R$.
 Choosing coordinates as in~\eqref{eq:i1MiuPB}, but indexing them with
 subscript to facilitate reading of following formulae, we have
\begin{equation*}
   {\mathcal K}^{-1} = \begin{pmatrix} 
                               K_{4\times 4}^{-1} \vline &  &  \cr \hline
                               \phantom{K_{4\times 4}^{-1}}  \vline & 0 & 1 \cr
                               \phantom{K_{4\times 4}^{-1}}  \vline & 1 & 0 \cr
                               \end{pmatrix} \,, \quad
{f \cdot \nu} =  \begin{pmatrix} 0 & 0 & 0 & 0 &
                     0 & 0 \cr 0 & 0 & 0 & 0 & -\nu_5 &
                     \nu_6 \cr 0 & 0 & 0 & 0 & 0 & 0 \cr 0 &
                      0 & 0 & 0 & 0 & 0 \cr 0 & \nu_5 & 0 & 0
                      & 0 &
                       \omega
                      \cr 0 & -\nu_6 & 0 & 0 & -\omega & 0 \cr
                 \end{pmatrix} \,,
\end{equation*}
 where $\omega=\left(\nu_1-2\nu_2+\nu_3+\nu_4\right)$. Notice, that
 quadratic Casimir of $su(2)$ constituent of $\ggg_0(\sw)$ read
\begin{equation}
  J_2 = \left(\nu_1-2\nu_2+\nu_3+\nu_4\right)^2 +  4 \nu_5 \nu_6 \,.
 \label{eq:SU2Casimir}
\end{equation}
  According to  \eqref{eq:AbelCoord} we introduce ``abelian''
  coordinates $\mu$:
\begin{equation}
\begin{split}
   \mu_i = \nu_i\,, \qquad &{} \qquad i=1,3,4 \,, \\
   \left(\mu_1 -2\mu_2 +\mu_3+\mu_4 \right)^2 &=
   \left(\nu_1-2\nu_2+\nu_3+\nu_4\right)^2 + 4 \nu_5 \nu_6  \,.
\end{split}
\label{eq:RiemSurf}
\end{equation}
 It is invariant with respect to $su(2)$ Weyl group, acting on $\mu$
 according to \eqref{eq:StandCox}:
\begin{equation}
   \mu_i \to \mu_i \,, \quad i=1,3,4 \,, \qquad \mu_2 \to  \mu_1+\mu_3+\mu_4 -
   \mu_2 \,. \label{eq:R2action}
\end{equation}
 Then group $R$ are generated by the following elementary reflections
 corresponding to three simple roots of $\sw$ degree one: $\alpha_1$,
 $\alpha_3$ and $\alpha_4$. Each corresponds gauge transformation
 \eqref{eq:FDgaugeact} with $n= x_1 E_{-\alpha_k} + x_2 E_{- \alpha_k
 + \alpha_2}$, with $k=1,3,4$. For each $x$ we find quadratic equation
 \eqref{eq:EqOnn} and thus we obtain, for instance, the following
 transformation for $\alpha_1$:
\begin{equation*}
\begin{split}
R_1^\pm : \nu_1 & \to  \frac{-{\nu_1} + {\nu_3} + {\nu_4} \pm
\sqrt{J_2} }{2} \,, \qquad \nu_3  \to {\nu_3} \,, \qquad \nu_4  \to {\nu_4} \,, \\
R_1^\pm: \nu_2 &  \to  \frac{-1} {2 \left(\omega + \nu_5 - \nu_6 \right) } 
\left( -2\nu_2 \nu_5-2 \left( 2\nu_2-2 \left( \nu_3 + \nu_4 \right)+ 
     \nu_5\right) \nu_6+ \right. \\ &\mspace{30mu}+
     \left(\nu_2 -2 \left( \nu_3+\nu_4\right)- 
     3\nu_6\right)\omega  +\left. {\omega }^2 \pm
       \sqrt{J_2} \left( \nu_1 + \nu_2 - \nu_3 - \nu_4 + \nu_6 \right)
         \right) , \\
R_1^\pm : \nu_5 & \to \frac{\left( 2 {\nu_1} - {\nu_2} + {\nu_5} \right)
     \left( \omega + 2\,{\nu_5} \mp 
       \sqrt{J_2} \right) }{2\,\left( \omega + {\nu_5} - {\nu_6}
\right) } \,, \\
R_1^\pm : \nu_6 & \to  \frac{\left( {\nu_1} + {\nu_2} - {\nu_3} - {\nu_4} + {\nu_6} \right) \,
     \left( \omega - 2 {\nu_6} \pm 
       \sqrt{J_2} \right) }{2 \left( \omega + {\nu_5} - {\nu_6}
\right) }  \,.
\end{split}
\end{equation*}
As illustration to Lemma \ref{lem:LinAction} note that 
\begin{equation*}
  J_2\left(R_1^\pm \left(\nu\right)\right) = \left( 
 \frac{3\mu_1-\mu_3-\mu_4 \pm \sqrt{J_2}}{2} \right)^2
\end{equation*}
 One can check that $R^\pm$ are not reflections, because, for example,
\begin{equation*}
  \left(R_1^+\right)^2 = \begin{cases} {\mathop{\rm Id_{Q_m^\ast}}} &
  \text{ if } \,  3\nu_1-\nu_3-\nu_4 \pm \sqrt{J_2} \geqslant 0 \,,\cr
  R_1^- & \text{ otherwise.}\end{cases}
\end{equation*}
 We can get rid of sign $\pm$, by making use of Weyl group $\Wgz$ of
 $sl(2)$ action \eqref{eq:R2action}. We define action of $R_1$ as
 $R_1^+$ for $\nu_2, \nu_5, \nu_6$ variables, choosing the following
 solution of eq.~\eqref{eq:RiemSurf}
\begin{equation*}
   \mu_2 = \frac{1}{2} \left\{ \sqrt{J_2} + \mu_1+ \mu_3 +\mu_4
   \right\} .
\end{equation*}
 This choice yields for $\mu$ coordinates
\begin{equation*}
   R_1 \colon \mu_{2} \to \mu_{2} \,, \quad \mu_{3} \to \mu_{3} \,,
   \quad \mu_{4} \to \mu_{4} \,, \quad \mu_1 \to - \mu_1
   + \mu_2 \,.
\end{equation*}
Then $R_1^- = R_2 R_1 R_2$, where $R_2$ acts on $\mu$ variables by
\eqref{eq:R2action} and acts trivially on $\nu_{2,5,6}$.  It can be
explicitly checked that so defined operation $R_1$ is a reflection
$R_1^2 ={\mathop{\rm Id}}$. It means that we have well defined
reflections on Riemann surface \eqref{eq:RiemSurf} over $Q_m^\ast$.

The same way we define reflections $R_3$ and $R_4$. Recall that
shortcut $\omega$ stands for $\omega=\mu_1 + \mu_3 + \mu_4 -2 \nu_2$  
\begin{equation*}
\begin{split}
R_3  \colon \mu_{1} &\to \mu_{1} \,, \quad \mu_2 \to \mu_2 \, \quad
   \mu_4 \to \mu_4 \, \quad \mu_3 \to - \mu_3 + \mu_2 \,, \\
R_3 : \nu_2 & \to \frac{1}{ 2 \left( \omega - \nu_5 + \nu_6 \right)} 
   \left[ 
2\nu_5\nu_6 +\nu_6\omega+\omega^2 - 2\mu_3 \left( 2\nu_6+\omega\right)+ 
\right. \\ &\mspace{30mu} + \left. \nu_2\left( -2\nu_5+4\nu_6+3\omega\right)+
{\sqrt{J_2}} \left( \mu_1 - \nu_2 - \mu_3 + \mu_4 + \nu_6 \right)\right] , \\
R_3 : \nu_5 & \to  \frac{\left( 2 \mu_3-\nu_2-\nu_5 \right) \,
     \left( -\omega  + 2 \nu_5+{\sqrt{J_2}}\right) }{2\left( \omega - 
       \nu_5 +\nu_6\right) } \,, \\
R_3 : \nu_6 & \to  \frac{\left( \mu_1-\nu_2-\mu_3+\mu_4+\nu_6\right) \,
     \left( \omega  + 2 \nu_6 + {\sqrt{J_2}} \right) }
     {2\left( \omega -\nu_5 + \nu_6 \right) } \,.
\end{split}
\end{equation*}
\begin{equation*}
\begin{split}
R_4  \colon \mu_{1} &\to \mu_{1} \,, \quad \mu_2 \to \mu_2 \,, \quad  
      \mu_4 \to \mu_4 \,, \quad \mu_3 \to - \mu_3+ \mu_2 \,, \\
R_4: \nu_2 & \to \frac{ \mu_1 + 2 \nu_2 + {\mu_3} - 3 \mu_4
+  {\sqrt{J_2}}}{2} \,, \qquad  \nu_6 \to {\nu_6} \,, \\ 
R_4 : \nu_5 & \to \frac{\left( -\nu_2 + 2 \mu_4 \right)
     \left( {\sqrt{J_2}} - \omega \right)}{2 \nu_6} \,.
\end{split}
\end{equation*}
Notice, that
$\mu_k=\nu_k$ for $k=1,3,4$ and we obtain that $R$ acts linearly on
these $\mu$. It is easy to check for these linear transformations, and
it certainly needs symbolic computation program to verify that
\begin{gather*}
  R_k^2=1\,, \qquad \left( R_1 R_3 \right)^2 = \left( R_1 R_4
  \right)^2 = \left( R_3 R_4 \right)^2 = 1\,, \\ \left( R_2 R_1
  \right)^3=\left( R_2 R_3 \right)^3=\left( R_2 R_4 \right)^3=1\,.
\end{gather*}
 That is they generate Weyl group of $D_4$ Lie algebra.

\paragraph{Solution to WDVV}
  By theorem~\ref{thm:solWDVV} we can extract a solution to WDVV from bi
Hamiltonian structure~\eqref{eq:solut}. Since scaling degree of fields
$\N_1,\dots,\N_4$ are $2,2,4,4$ respectively, we can find Euler vector
field
\begin{equation*}
  E = \left( \f{1}{2} \N_1, \f{1}{2} \N_2, \N_3, \N_3 \right) ,
\end{equation*}
 and the grade $d=1/2$.
 Following~\cite{DuFlat} Frobenius potential $F(\N)$ can be extracted
from~\eqref{eq:toextract}. It should be noted that
$\Gamma^{ij}_k = \partial_k \gamma^{j,i}$, making two relations equivalent.
We thus find the claimed free energy
\begin{equation}
\begin{split}
 \f{1}{4} F\left(\N\right) &= \N_2 \N_3 \N_4 + \f{1}{2} \N_1 \N_4^2 +
 \frac{\Del^5}{2^5 \cdot 3^4\cdot 5} + \f{1}{6} \N_1\N_3^2 -
 \f{1}{108} \N_3 \N_1^3 + \\ &+ \f{1}{12} \N_1\N_2^2 \N_3 + 
 \f{19}{2^8\cdot 3^4 \cdot5} \N_1^5 + 
 \f{7}{2^7\cdot 3^3} \N_1^3 \N_2^2 + \f{1}{3 \cdot 2^8} \N_1\N_2^4 \,.
\end{split}
\end{equation}
 It may be explicitly checked to verify WDVV
 equations~\eqref{eq:WDVV}.

%\begin{acknowledgement} 
\section*{Acknowledgement}
I would like to thank B.~Dubrovin for posing me this problem and for a
lot of valuable advice. I am indebted to A.~Maltsev for explaining me
his article and for enlightening discussions. I would like also to
thank G.~Falqui for stimulating talks.

 I would like to thank International School for Advanced Studies,
Italy, where this work has been done, for hospitality and creative
atmosphere.
%\end{acknowledgement}

\appendix
\section{Regular primitive conjugacy classes and their properties}
\label{app:uno}

 Here we collect information about regular primitive conjugacy classes
of Weyl group $\Wg$ for simple Lie algebras $\gg$. Classes are labeled
by the type of Coxeter diagram. Recall, that it coincides with the
Dynkin diagram for the Coxeter conjugacy class.
% \newlength{\mystuff}
%\setlength{\mystuff}{\abovedisplayskip}
%\setlength{\abovedisplayskip}{.54\mystuff}
%\setlength{\mystuff}{\belowdisplayskip}
%\setlength{\belowdisplayskip}{.54\mystuff}
\paragraph{$\mathbf{A_n}$}
\begin{eqnarray*}
 [w] &=& A_n \qquad  N_w=n+1 \qquad \sw=\s_p \\
% \sw &=& (1,1,\dots,1) \\
 I(w) &=& (1,2,3,\dots,n-1,n) = \Pr_w \qquad
 \ggg_0(\sw) = u(1)^{\otimes n}
\end{eqnarray*}
\paragraph{$\mathbf{B_n,C_n}$}
\begin{eqnarray*}
 [w] &=& B_n,C_n \qquad N_w = 2n \qquad \sw=\s_p \\
% \sw &=& (1,1,\dots,1) \\
 I(w) &=& (1,3,5\dots,2n-1) =  \Pr_w \qquad
 \ggg_0(\sw) = u(1)^{\otimes n}
\end{eqnarray*}
\paragraph{$\mathbf{D_n}$}
 We pick four last roots to form $D_4$ subalgebra.
\begin{eqnarray*}
  [w]&=&D_n \qquad N_w=2n-2 \qquad \sw=\s_p \\
%  \sw &=& (1,1,\dots,1) \\
  I(w) &=& (1,3,\dots,2n-1;n-1) = \Pr_w  \qquad
  \ggg_0(\sw) = u(1)^{\otimes n}
\end{eqnarray*}
\begin{eqnarray*}
 [w] &=& D_{2n}(a_{n-1}),D_{2n}(b_{n-1}) \qquad N_w=2n \\
 \sw &=& (1,\underbrace{1,0,1,0,\dots,1,0}_{2n-2 \text{ times}}, 1,1) \\
 I(w) &=& (\underbrace{1,1,3,3,\dots,2n-3,2n-3,2n-1,2n-1}_{2n \text{
 numbers}}) \\
 \Pr_w &=& (\underbrace{\overbrace{1,1,1,2},\dots,\overbrace{2n-3,2n-3,2n-3,2n-2}}_{n-1 \text{ groups in 4 elements}},2n-1,2n-1)  \\
 \ggg_0(\sw) &=& u(1)^{\otimes n+1} \oplus su(2)^{\otimes n-1}
\end{eqnarray*}
\paragraph{$\mathbf{G_2}$}
\begin{eqnarray*}
 [w] &=& G_2 \quad N_w = 6 \qquad \sw=\s_p \\
% \sw &=& (1,1,1) \\
 I(w) &=& (1,5) = \Pr_w \qquad \ggg_0(\sw) = u(1)^{\otimes 2}
\end{eqnarray*}
\paragraph{$\mathbf{F_4}$}
 $\alpha_1^2=\alpha^2_2=2$,$\alpha_3^2=\alpha_4^2=4$ with double bond between
the second and the third roots.
\begin{eqnarray*}
 [w] &=& F_4 \qquad N_w= 12 \qquad \sw=\s_p\\
% \sw &=& (1,1,1,1,1) \\
 I(w) &=& (1,5,7,11) = \Pr_w  \qquad  \ggg_0(\sw) = u(1)^{\otimes 4}
\end{eqnarray*}
\begin{eqnarray*}
 [w] &=& F_4(a_1) \qquad N_w=6 \\
 \sw &=& (1,1,0,1,0) \\
 I(w) &=& (1,1,5,5) \\
 \Pr_w &=& (1,1,1,2,3,4,5,5) \\
 \ggg_0(\sw) &=& u(1)^{\otimes 2} \oplus su(2)^{\otimes 2}
\end{eqnarray*}
\paragraph{$\mathbf{E_6}$}
 Roots $\alpha_6,\alpha_2,\alpha_3,\alpha_4$ form $D_4$ subalgebra.
\begin{eqnarray*}
 [w] &=& E_6 \qquad N_w=12 \qquad \sw=\s_p \\
% \sw &=& (1,1,1,1,1,1,1) \\
 I(w) &=& (1,4,5,7,8,11) = \Pr_w \qquad  \ggg_0(\sw) = u(1)^{\otimes 6}
\end{eqnarray*}
\begin{eqnarray*}
 [w] &=& E_6(a_1) \qquad N_w=9 \\
 \sw &=& (1,1,1,0,1,1,1) \\
 I(w) &=& (1,2,4,5,7,8) \\
 \Pr_w &=& (1,2,3,4,5,5,7,8) \\
 \ggg_0(\sw) &=& u(1)^{\otimes 5} \oplus su(2)
\end{eqnarray*}
\begin{eqnarray*}
 [w] &=& E_6(a_2) \qquad N_w=6 \\
 \sw &=& (1,1,0,1,0,1,0) \\
 I(w) &=& (1,1,2,4,5,5) \\
 \Pr_w &=& (1,1,1,2,2,2,3,3,4,4,5,5) \\
 \ggg_0(\sw) &=& u(1)^{\otimes 3} \oplus su(2)^{\otimes 3}
\end{eqnarray*}
\paragraph{$\mathbf{E_7}$}
\begin{eqnarray*}
 [w] &=& E_7 \qquad N_w = 18 \qquad \sw=\s_p \\
 I(w) &=& (1,5,7,9,11,13,17) = \Pr_w \qquad  \ggg_0(\sw)=u(1)^{\otimes 7}
\end{eqnarray*}
\begin{eqnarray*}
[w] &=& E_7(a_1)  \qquad N_w=14 \\
\sw &=& (1,1,1,0,1,1,1,1) \\
I(w) &=& (1,3,5,7,9,11,13) \\
\Pr_w &=& (1,3,5,5,7,8,9,11,13) \\
\ggg_0(\sw) &=& u(1)^{\otimes 6} \oplus su(2)
\end{eqnarray*}
\begin{eqnarray*}
 [w] &=& E_7(a_4) \qquad N_w=6 \\
 \sw &=& (1,0,0,1,0,0,1,0) \\
 I(w) &=& (1,1,1,3,5,5,5) \\
 \Pr_w &=&(1,1,1,1,1,1,2,2,2,2,3,3,3,3,3,4,4,4,5,5,5) \\
 \ggg_0(\sw) &=& u(1)^{\otimes 2} \oplus su(2) \oplus su(3)^{\otimes 2}
\end{eqnarray*}
\paragraph{$\mathbf{E_8}$}
\begin{eqnarray*}
 [w] &=& E_8 \qquad N_w=30 \qquad \sw=\s_p \\
 I(w) &=&(1,7,11,13,17,19,23,29) =\Pr_w \qquad
 \ggg_0(\sw) = u(1)^{\otimes 8}
\end{eqnarray*}
\begin{eqnarray*}
 [w] &=& E_8(a_1) \qquad N_w=24 \\
 \sw &=& (1,1,1,0,1,1,1,1,1) \\
 I(w) &=& (1,5,7,11,13,17,19,23) \\
 \Pr_w &=& (1,5,7,9,11,13,14,17,19,23) \\
 \ggg_0(\sw) &=& u(1)^{\otimes 7} \oplus su(2)
\end{eqnarray*}
\begin{eqnarray*}
 [w] &=& E_8(a_2) \qquad N_w=20 \\
 \sw &=& (1,1,1,0,1,0,1,1,1) \\
 I(w) &=& (1,3,7,9,11,13,17,19) \\
 \Pr_w &=& (1,3,5,7,8,9,11,11,13,14,17,19) \\
 \ggg_0(\sw) &=& u(1)^{\otimes 6} \oplus su(2)^{\otimes 2}
\end{eqnarray*}
\begin{eqnarray*}
 [w] &=& E_8(a_3),E_8(b_3) \qquad N_w=12 \\
 \sw &=& (1,1,0,1,0,0,1,0,0) \\
 I(w) &=& (1,1,5,5,7,7,11,11) \\
 \Pr_w &=& (1,1,1,2,3,4,5,5,5,5,6,6,7,7,7,8,9,10,11,11) \\
 \ggg_0(\sw) &=& u(1)^{\otimes 3} \oplus su(2)^{\otimes 3} \oplus su(3)
\end{eqnarray*}
\begin{eqnarray*}
 [w] &=& E_8(a_5),E_8(b_5) \qquad N_w=15 \\
 \sw &=& (1,1,0,1,0,1,0,1,0) \\
 I(w) &=& (1,2,4,7,8,11,13,14) \\
 \Pr_w &=& (1,2,3,4,5,5,7,7,7,8,9,9,11,11,13,14) \\
 \ggg_0(\sw) &=& u(1)^{\otimes 4} \oplus su(2)^{\otimes 4}
\end{eqnarray*}
\begin{eqnarray*}
 [w] &=& E_8(a_6) \qquad N_w=10 \\
 \sw &=& (1,0,0,1,0,0,1,0,0) \\
 I(w) &=& (1,1,3,3,7,7,9,9) \\
 \Pr_w &=& (1,1,1,2,3,3,3,3,3,4,4,4,5,5,5,6,6,6,7,7,7,8,9,9) \\
 \ggg_0(\sw) &=& u(1)^{\otimes 2} \oplus su(2)^{\otimes 2} \oplus su(3)^{\otimes 2}
\end{eqnarray*}
\begin{eqnarray*}
 [w] &=& E_8(a_8) \qquad N_w=6 \\
 \sw &=& (1,0,0,0,1,0,0,0,0) \\
 I(w) &=& (1,1,1,1,5,5,5,5) \\
 \Pr_w &=& (\underbrace{1,\dots,1}_{\text{10 times}},
            \underbrace{2,\dots,2}_{\text{10 times}},
            \underbrace{3,\dots,3}_{\text{10 times}},
            \underbrace{4,\dots,4}_{\text{6 times}},
            \underbrace{5,\dots,5}_{\text{4 times}}) \\
 \ggg_0(\sw) &=& u(1) \oplus su(4) \oplus su(5)
\end{eqnarray*}

%%%%%%%%%%%%%%%%%%%%%%%%%%%%%%%%%%%%%
%%%%%%%%%              %%%%%%%%%%%%%%
%%%%%%%%%  References  %%%%%%%%%%%%%%
%%%%%%%%%              %%%%%%%%%%%%%%
%%%%%%%%%%%%%%%%%%%%%%%%%%%%%%%%%%%%%

\end{document}